\documentclass[paper,notoc]{JHEP3}

\author{\footnote{KRYSTHAL collaboration} L.A. Harland-Lang$^1$, V.A. Khoze$^{1,2}$, M.G. Ryskin$^{2}$, W.J. Stirling$^{3}$\\ 
$^1$ Department of Physics and Institute for Particle Physics Phenomenology, \linebreak[4]University of Durham, DH1 3LE, UK\\  
$^2$ Petersburg Nuclear Physics Institute, NRC Kurchatov Institute, Gatchina, \linebreak[4]St. Petersburg, 188300, Russia\\ 
$^3$Cavendish Laboratory, University of Cambridge,
  J.J.\ Thomson Avenue, Cambridge, CB3 0HE, UK\\}

\title{Central exclusive production as a probe of the gluonic component of the $\eta'$ and $\eta$ mesons}

\abstract{Currently, the long--standing issue concerning the size of the gluonic content of the $\eta'$ and $\eta$ mesons remains unsettled. With this in mind we consider the central exclusive production (CEP) of $\eta'$, $\eta$ meson pairs in the perturbative regime, applying the Durham pQCD--based model of CEP and the `hard exclusive' formalism to evaluate the meson production subprocess. We calculate for the first time the relevant leading order parton--level processes $gg \to q\overline{q}gg$ and $gg \to gggg$, where the final--state $gg$ and $q\overline{q}$ pairs form a pseudoscalar flavour--singlet state. We observe that these amplitudes display some non--trivial and interesting theoretical properties, and we comment on their origin. Finally, we present a phenomenological study, and show that the cross sections for the CEP of $\eta'$, $\eta$ meson pairs are strongly sensitive to the size of the gluon content of these mesons. The observation of these processes could therefore provide 
important and novel insight into this problem.}

\preprint{IPPP/13/05\\  DCPT/13/10\\ Cavendish-HEP-13/01}

\usepackage{cite}
\usepackage{subfigure}
\usepackage{multirow}
\usepackage{helvet}
\usepackage{amsmath}
\usepackage{amssymb}
\usepackage{setspace}
\usepackage{setspace}
\usepackage[dvips]{graphicx}
\usepackage{epsfig}

\def\be{\begin{equation}}
\def\ee{\end{equation}}

\begin{document}

\section{Introduction}

Central exclusive production (CEP) processes of the type
\begin{equation}\label{exc}
pp({\bar p}) \to p+X+p({\bar p})\;,
\end{equation}
can significantly extend the physics programme at high energy hadron colliders. Here $X$ represents a system of invariant mass $M_X$, and the `$+$' signs denote the presence of large rapidity gaps. Such reactions provide a very promising way to investigate both QCD dynamics and new physics in hadron collisions, and consequently they have been widely discussed in the literature, with recently there being a renewal of interest in the CEP process, see for example~\cite{Martin:2009ku,Albrow:2010yb,HarlandLang:2013jf}  for reviews and further references.

It is well known that the $\eta$ and $\eta'$ mesons, the isoscalar members of the nonet of the lightest pseudoscalar mesons, play an important role in the understanding of low energy QCD. Knowledge of the quark and gluon components of their wave functions would provide important information about various aspects of non--perturbative QCD, see for instance~\cite{DiDonato:2011kr} and references therein. A clear observation of the presence of a purely gluonic, $gg$, component in the $\eta'$ (and $\eta$) mesons would also confirm that the gluons play an independent important role in hadronic spectroscopy. The presence of such a component of the $\eta'$ meson, due to the so called gluon anomaly~\cite{'tHooft:1976fv,Crewther:1977ce,Veneziano:1979ec,DiVecchia:1980ve}, is also related to the old question of why the $\eta'$ mass is much larger than that of the $\eta$ (the well--known `$U(1)$ problem', see~\cite{Weinberg:1975ui}).

Currently, while different determinations of the $\eta$--$\eta'$  mixing parameters are generally consistent, the long--standing issue concerning the extraction of the gluon content of the $\eta'$ (and $\eta$) remains uncertain, in particular due to non--trivial theory assumptions and approximations that must be made, as well as the current experimental uncertainties and limitations, see for example~\cite{Thomas:2007uy} for a discussion of the theoretical uncertainties (in e.g. the decay form factors) present in such extractions and~\cite{DiDonato:2011kr} for a review of the experimental situation. The results of a detailed analysis of various radiative processes and heavy particle decays, see for example~\cite{DiDonato:2011kr,Ricciardi:2012xu,Donskov:2013zp}, therefore do not currently allow a conclusive confirmation or otherwise of a non--$q\bar{q}$ component in the  the $\eta'$ and $\eta$ mesons, within the experimental uncertainties. Moreover, as discussed in~\cite{DiDonato:2011kr}, it is unlikely that lattice simulations of QCD will provide a determination of the gluonic contribution of the $\eta'$, $\eta$ wave functions in the near future. On the other 
hand, in~\cite{Kroll:2012hs} a study of the $\eta \gamma$ and $\eta' \gamma$ transition form factors, $F_{(\eta,\eta'),\gamma}(Q^2)$, appears to indicate that the two--gluon Fock component of the $\eta'$ meson may be quite large. 

In this paper we will show that the CEP of pseudoscalar meson pairs ($\eta'\eta'$, $\eta\eta'$, $\eta\eta$) at sufficiently high transverse momenta $p_\perp$ can provide a potentially powerful tool to probe  the structure of the  $\eta'$, $\eta$ mesons, and is especially well suited to addressing the old problem of the value of the gluonic flavour--singlet contribution, as well as clarifying the issue of $\eta$--$\eta'$ mixing. We show that any sizeable $gg$ component of the $\eta'$ (and $\eta$) can have a strong effect on the CEP cross section, and therefore such an exclusive process represents a sensitive probe of this. Moreover, we may expect that in the near future, after analysing the existing 4 photon candidates with $E_{T}>  2.5$ GeV and forward rapidity gaps, CDF will collect a large number of $\eta',\eta$ events and also significantly improve the current limits on $\pi^0\pi^0$ CEP~\cite{albrow}.  $\eta$ and $\eta'$ meson CEP may also be studied within the CMS/TOTEM special low--pileup runs with sufficient luminosity~\cite{albrow,MAdif12}. 

In~\cite{Aaltonen:2011hi} the observation of 43 $\gamma\gamma$ events with $|\eta(\gamma)|<1.0$ and  $E_T(\gamma)>2.5$ GeV, with no other particles detected in $-7.4<\eta<7.4$ was reported, which corresponds to a cross section of $\sigma_{\gamma\gamma}= 2.48^{+0.40}_{-0.35} $ $({\rm stat})^{+0.40}_{-0.51}$ $ ({\rm syst})$ pb. In~\cite{HarlandLang:2011qd} the $\pi^0\pi^0 \to 4\gamma$ CEP background was calculated for the first time and found to be small (with $\sigma(\pi^0\pi^0)/\sigma(\gamma\gamma)\sim 1\%$), a prediction that is in agreement with the CDF measurement, which finds that the contamination caused by $\pi^0\pi^0$   CEP is very small ($< 15$ events, corresponding to a ratio $N(\pi^0\pi^0)/N(\gamma\gamma)<0.35$, at 95\% C.L.). As well as representing a potential observable, we note that $\eta(')\eta(')$ CEP, via the $\eta(') \to \gamma\gamma$ decay may also in principle represent a background to $\gamma\gamma$ production, if this cannot be suppressed experimentally. It is therefore important to calculate the predicted $\eta(')\eta(')$ CEP cross sections, in particular in the presence of a potentially large $gg$ flavour--singlet component, which may enhance the corresponding rates.

We will apply the `hard exclusive' formalism described in~\cite{Brodsky:1981rp} (see also~\cite{Benayoun:1989ng}) to the production of a meson, $M$ anti--meson, $\overline{M}$, pair via the $gg \to M\overline{M}$ subprocess, which may then be used to calculate the corresponding CEP cross section, for $X=M\overline{M}$ in (\ref{exc}). In~\cite{HarlandLang:2011qd} the $gg \to q\overline{q}q\overline{q}$ parton--level amplitudes were calculated, where each $q\overline{q}$ pair forms a meson state, and from this the flavour singlet and non--singlet meson pair CEP cross section was calculated. However, any non--zero $gg$ component of the flavour singlet $\eta'$ and, through mixing, $\eta$ mesons will also be accessed via the $gg\to ggq\overline{q}$ and $gg \to gggg$ subprocesses. We will show that these amplitudes, which on the face of it are quite unrelated, in fact display a striking similarity, being identical with each other and with the `ladder--type' $gg \to q\overline{q} q\overline{q}$ amplitude calculated in~\cite{HarlandLang:2011qd}, that only contribute for flavour--singlet mesons, up to overall normalization factors. We show how this remarkable result may be explained in the MHV framework by the fact that the same external parton orderings contribute in all three cases.

Theoretical studies of meson pair CEP in fact have a long history, which predates the perturbative Durham approach depicted in Fig.~\ref{fig:pCp}. Exclusive $\pi\pi$ production, for example, mediated by Pomeron--Pomeron fusion, has been a subject of theoretical studies within a Regge--pole framework since the 1970s (see, for instance \cite{Pumplin:1976dm,Azimov:1974fa,Desai:1978rh} for early references and \cite{Lebiedowicz:2011nb,HarlandLang:2010ys,HarlandLang:2012qz} for more recent ones). However, as discussed in~\cite{HarlandLang:2010ys,HarlandLang:2012qz}, at comparatively large meson transverse momenta, $k_\perp$, CEP should be dominated by the perturbative 2--gluon exchange mechanism discussed above and shown in Fig.~\ref{fig:pCp}. At lower $k_\perp$ a study of the transition region between these `non--perturbative' and `perturbative' regimes may be necessary, as was performed in~\cite{HarlandLang:2012qz} for the case of $\pi\pi$ CEP. For the ($E_\perp<2.5$ GeV, $|\eta|<1$) meson pair event selection we will consider in this paper, the perturbative contribution was found to be dominant, and this is certainly expected to be true for the $\eta(')\eta(')$ perturbative cross sections, which we will show to be enhanced relative to $\pi^0\pi^0$ production. We will therefore neglect such a Regge--based non--perturbative contribution throughout this paper.

The outline of this paper proceeds as follows. In Section~\ref{CEPform} we introduce the CEP formalism for the process (\ref{exc}). In Section~\ref{fcalcsec} we introduce the `hard exclusive' formalism used to model the $gg \to M\overline{M}$ subprocess amplitudes, and calculate for the first time the $gg \to gg q\overline{q}$ and $gg \to gggg$ amplitudes through which the $gg$ component of the $\eta'$ and $\eta$ is accessed in the CEP process. In Appendix~\ref{mhvcalc} we show how these amplitudes may also be calculated using the MHV formalism, in a way which sheds some light on the interesting theoretical features which these amplitudes display. This aims to provide some theoretical insight into these amplitudes, but can be skipped by the reader who is only interested in the phenomenological implications of our analysis. In Section~\ref{res} we present numerical results for the CEP of $\eta'\eta'$, $\eta\eta'$ and $\eta\eta$ meson pairs, for a range of different sizes of the $gg$ component of 
the flavour--singlet distribution amplitude. Finally, we conclude in Section~\ref{conc}.

\section{Central exclusive production}\label{CEPform}

\begin{figure}[b]
\begin{center}
\includegraphics[scale=1.0]{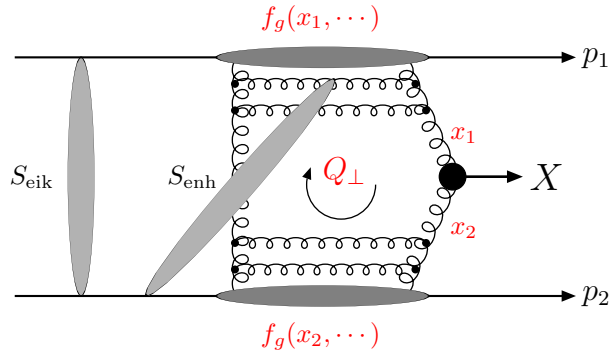}
\caption{The perturbative mechanism for the exclusive process $pp \to p\,+\, X \, +\, p$, with the eikonal and enhanced survival factors 
shown symbolically.}
\label{fig:pCp}
\end{center}
\end{figure} 

The formalism used to calculate the perturbative CEP cross section is explained in detail elsewhere~\cite{HarlandLang:2010ep} and so we will only review the relevant aspects here (for recent reviews and references see~\cite{Martin:2009ku,Albrow:2010yb,HarlandLang:2013jf}). The amplitude is described by the diagram shown in Fig.~\ref{fig:pCp}, where the hard subprocess $gg \to X$ is initiated by gluon--gluon fusion and the second $t$--channel gluon is needed to screen the colour flow across the rapidity gap intervals. We can write the `bare' amplitude in the factorized form~\cite{Khoze:2004yb,HarlandLang:2012qz}
\begin{equation}\label{bt}
T=\pi^2 \int \frac{d^2 {\bf Q}_\perp\, \mathcal{M}}{{\bf Q}_\perp^2 ({\bf Q}_\perp-{\bf p}_{1_\perp})^2({\bf Q}_\perp+{\bf p}_{2_\perp})^2}\,f_g(x_1,x_1', Q_1^2,\mu^2;t_1)f_g(x_2,x_2',Q_2^2,\mu^2;t_2) \; ,
\end{equation}
where the $f_g$'s in (\ref{bt}) are the skewed unintegrated gluon densities of the proton: in the kinematic region relevant to CEP, they are given in terms of the conventional (integrated) densities $g(x,Q_i^2)$. $t_i$ is the 4--momentum transfer squared to proton $i$ and $\mu$ is the hard scale of the process, taken typically to be of the order of the mass of the produced state: as in~\cite{Khoze:2004yb}, we use $\mu=M_X/2$ in what follows. The $t$--dependence of the $f_g$'s is isolated in a proton form factor, which we take to have the phenomenological form $F_N(t)=\exp(bt/2)$, with $b=4 \,{\rm GeV}^{-2}$. The $\mathcal{M}$ is the colour--averaged, normalised sub--amplitude for the $gg \to X$ process
\begin{equation}\label{Vnorm}
\mathcal{M}\equiv \frac{2}{M_X^2}\frac{1}{N_C^2-1}\sum_{a,b}\delta^{ab}q_{1_\perp}^\mu q_{2_\perp}^\nu V_{\mu\nu}^{ab} \; .
\end{equation}
Here $a$ and $b$ are colour indices, $M_X$ is the central object mass, $V_{\mu\nu}^{ab}$ represents the $gg \to X$ vertex and $q_{i_\perp}$ are the transverse momenta of the incoming gluons, given by
\begin{equation}
q_{1_\perp}=Q_\perp-p_{1_\perp}\,, \qquad
q_{2_\perp}=-Q_\perp-p_{2_\perp}\,,
\label{qperpdef}
\end{equation}
where $Q_\perp$ is the momentum transferred round the gluon loop and $p_{i_\perp}$ are the transverse momenta of the outgoing protons. Only one transverse momentum scale is taken into account in (\ref{bt}) by the prescription
\begin{align}\nonumber
Q_1 &= {\rm min} \{Q_\perp,|({\bf Q_\perp}-{\bf p}_{1_\perp})|\}\;,\\ \label{minpres}
Q_2 &= {\rm min} \{Q_\perp,|({\bf Q_\perp}+{\bf p}_{2_\perp})|\} \; .
\end{align}
The longitudinal momentum fractions carried by the gluons satisfy
\begin{equation}\label{xcomp}
\bigg(x' \sim \frac{Q_\perp}{\sqrt{s}}\bigg)  \ll \bigg(x \sim \frac{M_X}{\sqrt{s}}\bigg) \; ,
\end{equation} 
where $x'$ is the momentum fraction of the second $t$--channel gluon. The differential cross section at $X$ rapidity $y_X$ is then given by
\begin{equation}\label{ampnew}
\frac{{\rm d}\sigma}{{\rm d} y_X}=\langle S^2_{\rm enh}\rangle\int{\rm d}^2\mathbf{p}_{1_\perp} {\rm d}^2\mathbf{p}_{2_\perp} \frac{|T(\mathbf{p}_{1_\perp},\mathbf{p}_{2_\perp})|^2}{16^2 \pi^5} S_{\rm eik}^2(\mathbf{p}_{1_\perp},\mathbf{p}_{2_\perp})\; ,
\end{equation}
where $T$ is given by (\ref{bt}) and $S^2_{\rm eik}$ is the `eikonal' survival factor, the probability of producing no additional particles due to soft proton--proton rescattering. This is calculated using a generalisation of the `two--channel eikonal' model for the elastic $pp$ amplitude (see ~\cite{Khoze:2002nf} and references therein for details).

Besides the effect of eikonal screening, $S_{\rm eik}$, there is an additional suppression caused by the rescatterings of the intermediate partons (inside the unintegrated gluon distribution, $f_g$). This effect is described by the so--called enhanced Reggeon diagrams and usually denoted as $S^2_{\rm enh}$, see Fig.~\ref{fig:pCp}. The value of $S^2_{\rm enh}$ depends mainly on the transverse momentum of the corresponding partons, that is on the argument $Q^2_i$ of $f_g(x,x',Q^2_i,\mu^2;t)$ in (\ref{bt}), and depends only weakly on the $p_\perp$ of the outgoing protons (which formally enters only at NLO). While in~\cite{HarlandLang:2010ep,HarlandLang:2009qe} the $S^2_{\rm enh}$--factor was calculated using the formalism of~\cite{Ryskin:2009tk}, here, following~\cite{HarlandLang:2011qd,HarlandLang:2012qz}, we use a newer version of the multi--Pomeron model~\cite{Ryskin:2011qe} which incorporates the continuous dependence on $Q^2_i$ and not only three `Pomeron components' with different `mean' $Q_i$. We 
therefore include the $S_{\rm enh}$ factor inside the integral (\ref{bt}), with $\langle S^2_{\rm enh}\rangle$ being its average value integrated over $Q_\perp$.

If we consider the exact limit of forward outgoing protons, $p_{i_\perp}=0$, then we find that after the $Q_\perp$ integration (\ref{Vnorm}) reduces to
\begin{equation}\label{mprop}
\mathcal{M}\propto q_{1_\perp}^i q_{2_\perp}^j V_{ij} \to \frac{1}{2}Q_\perp^2(V_{++}+ V_{--})\sim\sum_{\lambda_1,\lambda_2}\delta^{\lambda_1\lambda_2}V_{\lambda_1\lambda_2}\;,
\end{equation}
where $\lambda_{(1,2)}$ are the gluon helicities in the $gg$ rest frame. The only contributing helicity amplitudes are therefore those for which the $gg$ system is in a $J_z=0$ state, where the $z$--axis is defined by the direction of motion of the gluons in the $gg$ rest frame, which, up to corrections of order $\sim q_\perp^2/M_X^2$, is aligned with the beam axis.  In general, the outgoing protons can pick up a small $p_\perp$, but large values are strongly suppressed by the proton form factor, and so the production of states with non--$J_z=0$ quantum numbers is correspondingly suppressed (see~\cite{HarlandLang:2010ep,HarlandLang:2009qe} for examples of this in the case of $\chi_{(c,b)}$ and $\eta_{(c,b)}$ CEP). In particular, we find roughly that
\begin{equation}\label{simjz2}
\frac{|T(|J_z|=2)|^2}{|T(J_z=0)|^2} \sim \frac{\langle p_\perp^2 \rangle^2}{\langle Q_\perp^2\rangle^2}\;,
\end{equation}
which is typically of order $\sim1/50-1/100$, depending on the central object mass, cms energy $\sqrt{s}$ and choice of PDF set. As discussed in~\cite{HarlandLang:2011qd}, this `$J_z=0$ selection rule'~\cite{Kaidalov:2003fw} will have important consequences for the case of meson pair CEP. Finally, we note that in (\ref{mprop}) the incoming gluon helicities are averaged over at the {\it amplitude} level: this result is in complete contrast to a standard inclusive production process where the {\it amplitude squared} is averaged over all gluon helicities. Eq. (\ref{mprop}) can be readily generalised to the case of non--$J_z=0$ gluons which occurs away from the forward proton limit, see in particular Section 4.1 (Eq. (41)) of~\cite{HarlandLang:2010ep}, which we make use of throughout to calculate the $M\overline{M}$ CEP amplitude from the corresponding $gg\to M\overline{M}$ helicity amplitude.

\section{$gg \to \eta(')\eta(')$ amplitudes: Feynman diagram calculation}\label{fcalcsec}

\subsection{The hard exclusive formalism}

\begin{figure}
\begin{center}
\subfigure[]{\includegraphics[scale=1.2]{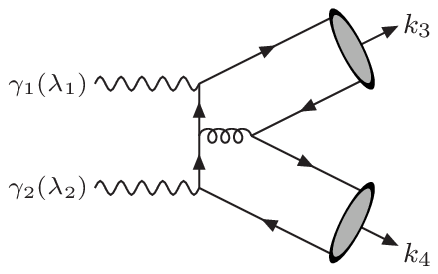}}
\subfigure[]{\includegraphics[scale=1.2]{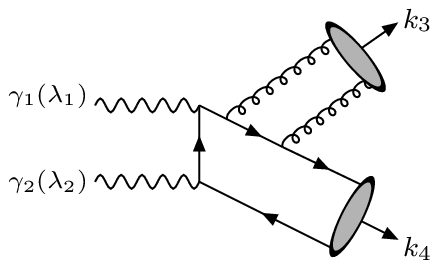}}
\caption{Representative Feynman diagrams for the $\gamma\gamma \to q\overline{q}q\overline{q}(gg)$ processes. There are 20 Feynman diagrams of type (a), and the corresponding helicity amplitudes are given in~\cite{HarlandLang:2011qd} (see also ~\cite{Benayoun:1989ng,Brodsky:1981rp}). There are 24 diagrams of type (b), and the corresponding helicity amplitudes are given by (\ref{tggam0}, \ref{tggam2}). There are 31 diagrams of type (a), but with the photons replaced by gluons, and the corresponding helicity amplitudes are given in~\cite{HarlandLang:2011qd}.}\label{feyn1}
\end{center}
\end{figure}

The leading order contributions to the $\gamma\gamma \to M\overline{M}$ process were first calculated in~\cite{Brodsky:1981rp} (see also~\cite{Benayoun:1989ng,Chernyak:2006dk,Chernyak:2012pw}), where $M(\overline{M})$ is in this case a flavour non--singlet meson(anti--meson). The cross section has been calculated at NLO in~\cite{Duplancic:2006nv}. For the case of mesons with flavor--singlet Fock states there is also a contribution coming from the LO two--gluon component of the meson, and this was calculated in~\cite{Atkinson:1983yh} (see also~\cite{Wakely:1991ej,Baier:1985wv}). In particular, in~\cite{Atkinson:1983yh} they considered the process $\gamma\gamma \to \eta_1 M$ (where $M\neq \eta_1$), where $\eta_1$ is a flavour--singlet meson state that can has both a $q\overline{q}$ and a $gg$ component. The amplitude can be written as
\begin{equation}\label{amp}
\mathcal{M}_{\lambda\lambda'}(\hat{s},\theta)=\int_{0}^{1} \,{\rm d}x \,{\rm d}y\, \phi_{1}(x)\phi_M(y)\, T_{\lambda\lambda'}^q(x,y;\hat{s},\theta)
+\int_{0}^{1} \,{\rm d}x \,{\rm d}y\, \phi_{G}(x)\phi_M(y)\, T_{\lambda\lambda'}^g(x,y;\hat{s},\theta)\;,
\end{equation}
where $\hat{s}$ is the $\eta_1 M$ invariant mass, $x,y$ are the meson momentum fractions carried by the quarks or gluons in the meson, $\lambda$, $\lambda'$ are the photon helicities and $\theta$ is the scattering angle in the $\gamma\gamma$ cms frame. $T_{\lambda\lambda'}^{q(g)}$ is the hard scattering amplitude for the parton level process $\gamma\gamma\to q\overline{q}\,q\overline{q}(gg)$, see Fig.~\ref{feyn1}, where each $q\overline{q}$ and $gg$ pair is collinear and has the appropriate colour, spin, and flavour content projected out to form the parent meson. In the meson rest frame, the relative motion of the partons is small: thus for a meson produced with large momentum, $|\vec{k}|$, we can neglect the transverse component of the parton momentum, $\vec{q}$, with respect to $\vec{k}$, and simply write $q=xk$ in the calculation of $T_{\lambda\lambda'}$. $\phi(x)$ is the leading twist meson distribution amplitude, representing the probability amplitude of finding a valence parton in the meson carrying a longitudinal momentum fraction $x$ of the meson's momentum, integrated up to the scale $Q$ over the quark transverse momentum $\vec{q_t}$ (with respect to meson momentum $\vec k$). While $\phi_M(x)$ and $\phi_{1}(y)$ represent the $q\overline{q}$ distribution amplitudes of the mesons $M$ and $\eta_1$, respectively, $\phi_{G}$ corresponds to the $gg$ distribution amplitude of the $\eta_1$.

We recall (see~\cite{HarlandLang:2011qd} and references therein for further details) that the meson distribution amplitude depends on the (non--perturbative) details of hadronic binding and cannot be predicted in perturbation theory. However, the overall normalization of the $q\overline{q}$ distribution amplitude can be set by the meson decay constant $f_M$ via~\cite{Brodsky:1981rp}\footnote{We note that this normalization, which we use throughout this paper, corresponds to the definition of $f_M$ given in~\cite{Brodsky:1981rp} for the case of the pion, that is it with $f_\pi \approx 93$ MeV. In the literature the value $f_\pi\to  \sqrt{2} f_\pi\approx 133$ MeV, is often used, in particular in~\cite{Kroll:2002nt,Kroll:2012hs}, which we will refer to later. In Section 5 of~\cite{HarlandLang:2011qd}, the more conventional value $f_\pi \approx 133$ MeV is confusingly quoted, but in all numerical calculations the lower value was correctly taken.}
\begin{equation}\label{wnorm}
\int_{0}^{1}\, {\rm d}x\,\phi_M(x)=\frac{f_M}{2\sqrt{3}}\;.
\end{equation}
It was also shown in~\cite{Lepage:1980fj} that for very large $Q^2$ the meson distribution amplitude evolves towards the asymptotic form
\begin{equation}\label{asym}
\phi_M(x,Q)\underset{Q^2\to \infty}{\to} \,\sqrt{3} f_M\, x(1-x)\;.
\end{equation}
However this logarithmic evolution is very slow and at realistic $Q^2$ values the form of $\phi_M$ can in general be quite different. Indeed, the recent BABAR  measurement of the pion transition form factor $F_{\pi\gamma}(Q^2)$~\cite{Aubert:2009mc,Druzhinin:2009gq}, for example, strongly suggests that $\phi_\pi(x,Q)$ does not have the asymptotic form out to $Q^2\lesssim 40\,{\rm GeV}^2$, although new Belle data~\cite{Uehara:2012ag} are in conflict with this. Another possible choice is the `Chernyak--Zhitnisky' (CZ) form, which we will make use of later on~\cite{Chernyak:1981zz}
\begin{equation}\label{CZ}
\phi^{{\rm CZ}}_M(x,Q^2=\mu_0^2)=5\sqrt{3}f_M\, x(1-x)(2x-1)^2\;,
\end{equation}
where the starting scale is roughly $\mu_0\approx 1$ GeV. For the two--gluon distribution amplitude, $\phi_G(x)$, the normalization cannot be set as in (\ref{wnorm}), as we have $\phi_G(x)=-\phi_G(1-x)$, as required by the antisymmetry of the pseudoscalar spin projection (\ref{cgg}) of the gluons under this interchange, but an analogous formula can be written down~\cite{Ohrndorf:1981uz,Atkinson:1983yh,Baier:1985wv}
\begin{equation}
\int_0^1 {\rm d}y\, \phi_G(x,Q^2) (2x-1) \propto f_G(Q^2)\;.
\end{equation}
Such an expression serves to define $f_G$, the value of which is to be determined. While as $Q^2\to \infty$, it can be shown that the $gg$ distribution amplitude vanishes due to QCD evolution~\cite{Ohrndorf:1981uz,Baier:1981pm}
\begin{equation}
\lim_{Q^2 \to \infty} \phi_G(x)=0\;,
\end{equation}
there is no reason to assume this will be the case at experimentally relevant energies.

More precisely, the $q\overline{q}$ flavour--singlet and $gg$ distribution amplitudes can be expanded in terms of the Gegenbauer polynomials $C_n$~\cite{Lepage:1980fj,Ohrndorf:1981uz,Baier:1981pm}
\begin{align} \nonumber
\phi_{(1,8),M}(x,\mu_F^2)&=\frac{6 f_{(1,8)}^M}{2\sqrt{N_C}} x(1-x)[1+\sum_{n=2,4,\cdots} a_n^{(1,8)}(\mu_F^2)C_n^{3/2}(2x-1)]\;,\\ \label{waves}
\phi_{G,M}(x,\mu_F^2)&=\frac{f_1^M}{2\sqrt{N_C}}\sqrt{\frac{C_F}{2n_f}} x(1-x)\sum_{n=2,4,\cdots} a_n^G(\mu_F^2) C_{n-1}^{5/2}(2x-1)\;,
\end{align}
where $\mu_F$ is the factorization scale, taken as usual to be of the order of the hard scale of the process being considered, and $n_f=3$ for $\eta(')$ mesons. The $f_{1,8}^M$ (with $M=\eta,\eta'$ in the present case) are given by (\ref{etafit}), with the $M$ dependence expressing the difference due to the mixing of the $\eta$, $\eta'$ states and decay constants. We assume that apart from this the distribution amplitudes, i.e. the $a_n$, are independent of the meson being considered. We choose to include the decay constants $f_1^M$, $f_8^M$ explicitly in our definition of the distribution amplitudes, in contrast to, e.g.~\cite{Kroll:2002nt,Kroll:2012hs}, where they are introduced in the hard amplitude $T_{\lambda\lambda'}$ in (\ref{amp})\footnote{Our choice of normalization of $\phi_{1,G}$ differs further by a factor of $1/2\sqrt{N_C}$, which in~\cite{Kroll:2012hs} is included in the hard amplitude, and an additional factor of $\sqrt{C_F/2n_f}x(1-x)$ in $\phi_G$ due to the different normalization of the $gg$ spin projection (\ref{cgg}). Of course, the final physical result will not depend on the choice of convention, which simply corresponds to a choice of which overall factors to include in the distribution amplitude $\phi_{G,1}$, and which to include in the hard amplitude $T_{\lambda\lambda'}$ in (\ref{amp}).}.

The evolution of the distribution amplitude is then dictated by the $\mu_F^2$ dependence of the coefficients $a_n$ via
\begin{align}\nonumber
a_n^1(\mu_F^2)&=a_n^{(+)}(\mu_0^2)\left(\frac{\alpha_s(\mu_0^2)}{\alpha_s(\mu_F^2)}\right)^{\gamma_n^{(+)}/\beta_0}+\rho^{(-)}_n a_n^{(-)}(\mu_0^2)\left(\frac{\alpha_s(\mu_0^2)}{\alpha_s(\mu_F^2)}\right)^{\gamma_n^{(-)}/\beta_0}\;,\\ \label{b2gev}
a_n^G(\mu_F^2)&=\rho_n^{(+)}a_n^{(+)}(\mu_0^2)\left(\frac{\alpha_s(\mu_0^2)}{\alpha_s(\mu_F^2)}\right)^{\gamma_n^{(+)}/\beta_0}+ a_n^{(-)}(\mu_0^2)\left(\frac{\alpha_s(\mu_0^2)}{\alpha_s(\mu_F^2)}\right)^{\gamma_n^{(-)}/\beta_0}\;.
\end{align}
That is, the quark and gluon components mix under evolution. Here \linebreak[4]$\beta_0=11-2n_f/3$, the $a_n^{(\pm)}(\mu_0^2)$ at the starting scale $\mu_0$ are inputs which must be, e.g., extracted from data, and the remaining factors $\gamma^\pm$, $\rho^{(\pm)}$ are defined in Appendix~\ref{ap1}. We note that despite the different normalization convention we take in (\ref{waves}) for the quark and gluon distribution amplitudes, the coefficients $a_n$ are defined so that they obey the same evolution equation as in for example~\cite{Kroll:2012hs}. While the size of the $a_n^G(\mu_F^2=Q^2)$ will decrease with increasing scale, their initial value can in principle be quite large, and the evolution of for example the leading coefficient $a_2^G \to 0$ with $Q^2$ is slow\footnote{The higher order $n=4,6,$... terms are found to evolve faster towards zero with increasing $n$. Combined with the fact that as $n$ increases, the additional powers of $C_{n-1}^{5/2}(2x-1)$ (or $C_n^{3/2}(2x-1)$ in the $q\overline{q}$ case) give a smaller numerical contribution to the distribution amplitude, this means that any fit can effectively truncate the series (\ref{waves}) after a limited number of terms.}.

Finally, to make contact with the physical $\eta$, $\eta'$ states we will be considering in this paper, we introduce the flavour--singlet and non--singlet quark basis states
\begin{align} \nonumber
|q\overline{q}_1\rangle &=\frac{1}{\sqrt{3}}|u\overline{u}+d\overline{d}+s\overline{s}\rangle\;,\\ \label{qfock}
|q\overline{q}_8\rangle &=\frac{1}{\sqrt{6}}|u\overline{u}+d\overline{d}-2s\overline{s}\rangle\;,
\end{align}
and the two--gluon state
\begin{equation}\label{gfock}
|gg\rangle\;,
\end{equation}
with corresponding distribution amplitudes given by (\ref{waves}). Here, we follow~\cite{Feldmann:1997vc} (see also~\cite{Kiselev:1992ms,Leutwyler:1997yr}) in taking a general two--angle mixing scheme for the $\eta$ and $\eta'$ mesons. That is, the mixing of the $\eta$, $\eta'$ decay constants is not assumed to follow the usual (one--angle) mixing of the states. This is most easily expressed in terms of the $\eta$ and $\eta'$ decay constants
\begin{align}\nonumber
f_8^\eta=f_8 \cos \theta_8\;,   \qquad\qquad f_1^\eta&=-f_1 \sin \theta_1 \;,\\ \label{etafit}
f_8^{\eta'}=f_8 \sin \theta_8\;,     \qquad\qquad f_1^{\eta'}&=f_1 \cos \theta_1 \;,
\end{align}
with the fit of~\cite{Feldmann:1998vh} giving
\begin{align}\nonumber
  f_8=1.26 f_\pi\;, \qquad & \qquad \theta_8   = -21.2^\circ\;,\\ \label{thetafit}
 f_1=1.17 f_\pi\;,  \qquad & \qquad \theta_1  = -9.2^\circ\;.
\end{align}
We then take the distribution amplitudes (\ref{waves}) with the decay constants given as in (\ref{etafit}), for the corresponding Fock components (\ref{qfock}) and (\ref{gfock}). That is, the $\eta$ and $\eta'$ states are given schematically by (see for example~\cite{Kroll:2002nt,Kroll:2012hs} for more details and discussion)
\begin{align} \nonumber
|\eta\rangle &= f_8\cos \theta_8 \left[\tilde{\phi}_{8,\eta}(x,\mu_F^2)|q\overline{q}_8\rangle \right]-f_1\sin \theta_1 \left[\tilde{\phi}_{1,\eta}(x,\mu_F^2)|q\overline{q}_1\rangle+\tilde{\phi}_{G,\eta}(x,\mu_F^2)|gg\rangle\right]\;,\\ \label{mixing}
|\eta'\rangle &= f_8\sin \theta_8 \left[\tilde{\phi}_{8,\eta'}(x,\mu_F^2)|q\overline{q}_8\rangle\right] +f_1\cos \theta_1 \left[\tilde{\phi}_{1,\eta'}(x,\mu_F^2)|q\overline{q}_1\rangle+\tilde{\phi}_{G,\eta'}(x,\mu_F^2)|gg\rangle\right]\;,
\end{align}
where to make things explicit the distribution amplitudes $\tilde{\phi}$ are defined as in (\ref{waves}), but with the decay constants divided out (i.e. $\tilde{\phi}_{8,\eta}(x,\mu_F^2)=\phi_{8,\eta}(x,\mu_F^2)/f_8^\eta$...), and these do not represent the conventional, normalized expressions for the $\eta',\eta$ Fock states, but simply indicate the decay constants and distribution amplitudes that should be associated with each $q\overline{q}$ and $gg$ state in this two--angle mixing scheme.

\subsection{Preliminary consideration: $\gamma \gamma \to M \eta_1$}\label{prelim}

We begin this section by recalculating the $\gamma\gamma \to \eta_1 M$ helicity amplitudes, as in~\cite{Atkinson:1983yh}. For simplicity, we consider the case of a pure flavour--singlet state $\eta_1$, but the following result can readilty be applied to the more realistic case of an $\eta'$ or $\eta$ meson. The appropriate $qq$ and $gg$ spin and colour quantum numbers are projected onto the relevant pseudoscalar meson states using~\cite{Lepage:1980fj,Kroll:2002nt}
\begin{align}\label{pqq}
\mathcal{P}_{q\overline{q}}&=\frac{\delta_{ij}}{\sqrt{N_C}}\left(\frac{\gamma_5{\not}p_M}{\sqrt{2}}\right)\;,\\ \label{pgg}
\mathcal{P}_{gg}&=-\frac{\delta_{ab}}{\sqrt{N_C^2-1}}\frac{i\epsilon_\perp^{\mu\nu}}{\sqrt{2}}\;,
\end{align}
where $\epsilon_\perp^{12}=-\epsilon_\perp^{21}=1$ and all other components are zero. It can be expressed as
\begin{equation}
\epsilon_\perp^{\mu\nu}=\frac{2}{\hat{s}}e^{\mu\nu\alpha\beta}p^M_\alpha p^{\eta_1}_\beta\;,
\end{equation}
where $\epsilon^{0123}=-1$. Explicitly, these correspond to the following combinations of spin states
\begin{align}\label{cpp}
\frac{1}{\sqrt{2}}\left(\frac{u_+(y)}{\sqrt{y}}\frac{\overline{v}_-(1-y)}{\sqrt{1-y}}+\frac{u_-(y)}{\sqrt{y}}\frac{\overline{v}_+(1-y)}{\sqrt{1-y}}\right)&=\frac{\gamma_5 {\not}p_M}{\sqrt{2}}\;,\\ \label{cgg}
\frac{1}{\sqrt{2}}\left(\epsilon^{*\mu}_+(x)\epsilon^{*\nu}_-(1-x)-\epsilon^{*\mu}_-(x)\epsilon^{*\nu}_+(1-x)\right)&=-i\frac{\epsilon^{\mu\nu}_\perp}{\sqrt{2}}\;,
\end{align}
where $u(z),v(z)$ are the usual Dirac spinors for the quarks and $\epsilon(z)$ is the polarization vector of the gluon, carrying momentum fraction $z$ of the parent mesons, while the $\pm$ signs indicate the particle helicities. The normalizations are conventional (other choices are possible, provided the distribution amplitude is suitably re--defined to compensate), and we can see that these projections are odd under the parity transformation $+ \leftrightarrow -$, as required for a $P=-1$ meson state. We make use of (\ref{pqq}, \ref{pgg}) throughout this paper. 


Explicitly calculating the amplitudes corresponding to the 24 contributing Feynman diagrams (see e.g. Fig.~1 (c) of~\cite{Atkinson:1983yh})\footnote{We have made use of the FORM symbolic manipulation programme~\cite{Vermaseren:2000nd} throughout this paper.}, we find
\begin{align} \label{tggam0}
T^{g(\gamma\gamma)}_{++}&=T^{g(\gamma\gamma)}_{--}=0\;,\\ \label{tggam2}
T^{g(\gamma\gamma)}_{+-}&=T^{g(\gamma\gamma)}_{-+}\stackrel{!}{=}\sqrt{\frac{N_C^2-1}{N_C}}\frac{64\pi^2 \alpha e_q^2\alpha_s}{\hat{s}y(1-y)}\frac{b\cos^2\theta-(2x-1)a}{a^2-b^2\cos^2\theta}\;,
\end{align}
where the $(\gamma\gamma)$ is to distinguish these from the amplitudes with initial--state gluons that we will consider shortly. $e_q$ is the quark charge in the meson $M$  and $a,b$ are given by
\begin{align}\label{a}
a&=(1-x)(1-y)+xy\; ,\\ \label{b}
b&=(1-x)(1-y)-xy\; .
\end{align}
The `!' in (\ref{tggam2}) indicates that this is not an exact equality, but rather the result of the explicit calculation and (\ref{tggam2}) are equivalent {\it after} the amplitude $T^{g(\gamma\gamma)}$ has been integrated over the (antisymmetric) $\eta_1$ distribution amplitude $\phi_{G,\eta_1}(x)$. That is, they are equivalent up to terms which are even under the interchange $x \to 1-x$. We note that (\ref{tggam0}) is consistent with the results of~\cite{Atkinson:1983yh,Wakely:1991ej}, while (\ref{tggam2}) is consistent up to overall numerical factors (in particular our result is a factor of 2 larger than that of~\cite{Wakely:1991ej}). As no definition of the gluon spin projector (\ref{pgg}) is given in~\cite{Atkinson:1983yh,Wakely:1991ej}, while the results also differ between these two papers by a factor of $2x(1-x)$, a precise comparison is quite difficult to make, and we can reasonably assume that the difference between (\ref{tggam2}) and the results~\cite{Atkinson:1983yh,Wakely:1991ej} is due to differing normalization conventions for the spin projectors and gluon wavefunctions.

\subsection{$gg \to gg q\overline{q}(gg)$ amplitudes}\label{explicit}

\begin{figure}
\begin{center}
\subfigure[]{\includegraphics[scale=0.75]{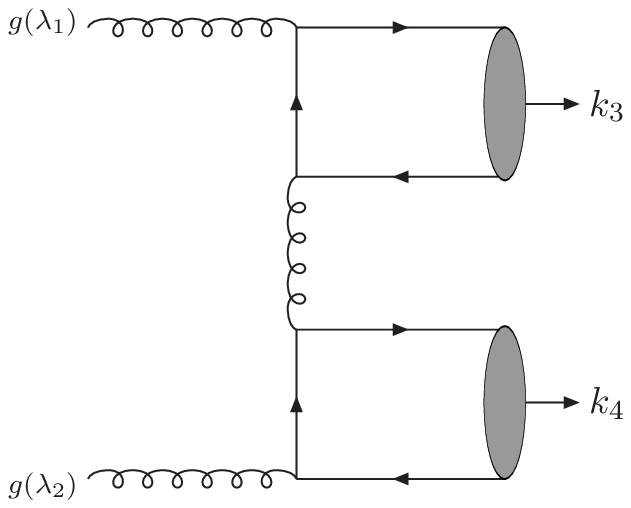}}
\subfigure[]{\includegraphics[scale=0.75]{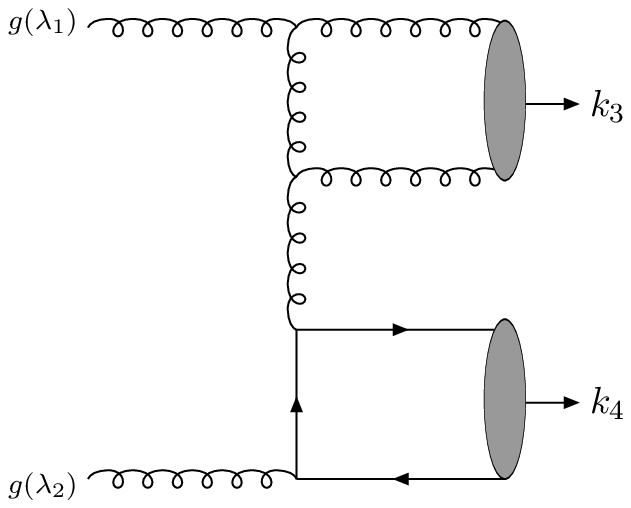}}
\subfigure[]{\includegraphics[scale=0.75]{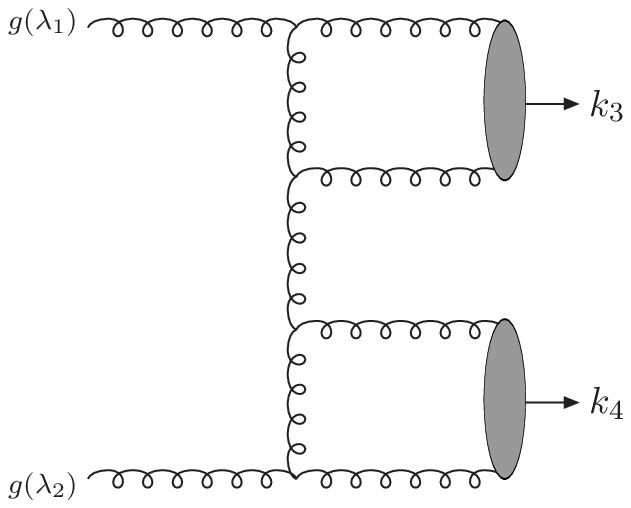}}
\caption{Representative Feynman diagrams for the $gg \to M\overline{M}$ process, where the $M$ are flavour--singlet mesons. There are 8 Feynman diagrams of type (a), and the corresponding helicity amplitudes are given by (\ref{lad0}, \ref{lad2}). There are 76 Feynman diagrams of type (b), and the corresponding $J_z=0$ helicity amplitude is given by (\ref{tgq0}), while a numerical evaluation of the $|J_z|=2$ case is shown in Fig.~\ref{ampplots}. There are 130 Feynman diagrams of type (c), and the corresponding $J_z=0$ helicity amplitude is given by (\ref{tgg0}), while a numerical evaluation of the $|J_z|=2$ case is shown in Fig.~\ref{ampplots}. In the case of the amplitudes for (b) and (c), all diagrams allowed by colour conservation are included, and not just diagrams of this ladder type.}\label{feyn2}
\end{center}
\end{figure}

Representative diagrams for the three processes we will consider are shown in Fig.~\ref{feyn2}: in the case of the $4g$ and $6g$ amplitudes, we note that all diagrams allowed by colour conservation and the antisymmetry of the gluon spin projection contribute to the total amplitude, and not just diagrams of this ladder type.
After a quite lengthy calculation, we find that the $gg \to ggq\overline{q}$ amplitude for $J_z=0$ incoming gluons is given by
\begin{equation}\label{tgq0}
 T^{gq}_{++}=T^{gq}_{--}=-2\,\delta^{ab}\sqrt{\frac{N_C}{N_C^2-1}}\frac{64\pi^2 \alpha_s^2}{\hat{s}xy(1-x)(1-y)}\frac{(1+\cos^2\theta)}{(1-\cos^2\theta)^2}(2x-1)\;,
\end{equation}
The $gg \to gggg$ amplitude for $J_z=0$ incoming gluons is given by
\begin{equation}\label{tgg0}
T^{gg}_{++}=T^{gg}_{--}=-4\,\delta^{ab}\frac{N_C^2}{N_C^2-1}\frac{64\pi^2 \alpha_s^2}{\hat{s}xy(1-x)(1-y)}\frac{(1+\cos^2\theta)}{(1-\cos^2\theta)^2}(2x-1)(2y-1)\;,
\end{equation}
We also reproduce for completeness the $gg\to q\overline{q}q\overline{q}$ amplitudes for flavour--singlet mesons corresponding to the `ladder diagrams' as in Fig.~\ref{feyn2} (a), see~\cite{HarlandLang:2011qd}.
\begin{align}\label{lad0}
T_{++}^{qq.}=T_{--}^{qq.}&=-\frac{\delta^{ab}}{N_C}\frac{64\pi^2 \alpha_S^2}{\hat{s}xy(1-x)(1-y)}\frac{(1+\cos^2 \theta)}{(1-\cos^2 \theta)^2}\;,\\ \label{lad2}
T_{+-}^{qq.}=T_{-+}^{qq.}&=-\frac{\delta^{ab}}{N_C}\frac{64\pi^2 \alpha_S^2}{\hat{s}xy(1-x)(1-y)}\frac{(1+3\cos^2 \theta)}{2(1-\cos^2 \theta)^2}\;.
\end{align}
Thus all three $J_z=0$ amplitudes are identical, up to overall colour and normalization factors (including the factors of `$(1-2x)$, $(1-2y)$' which ensure that the amplitudes have the correct symmetry under the interchange $x(y)\leftrightarrow 1-x(y)$).

\begin{figure}
\begin{center}
\includegraphics[scale=0.6]{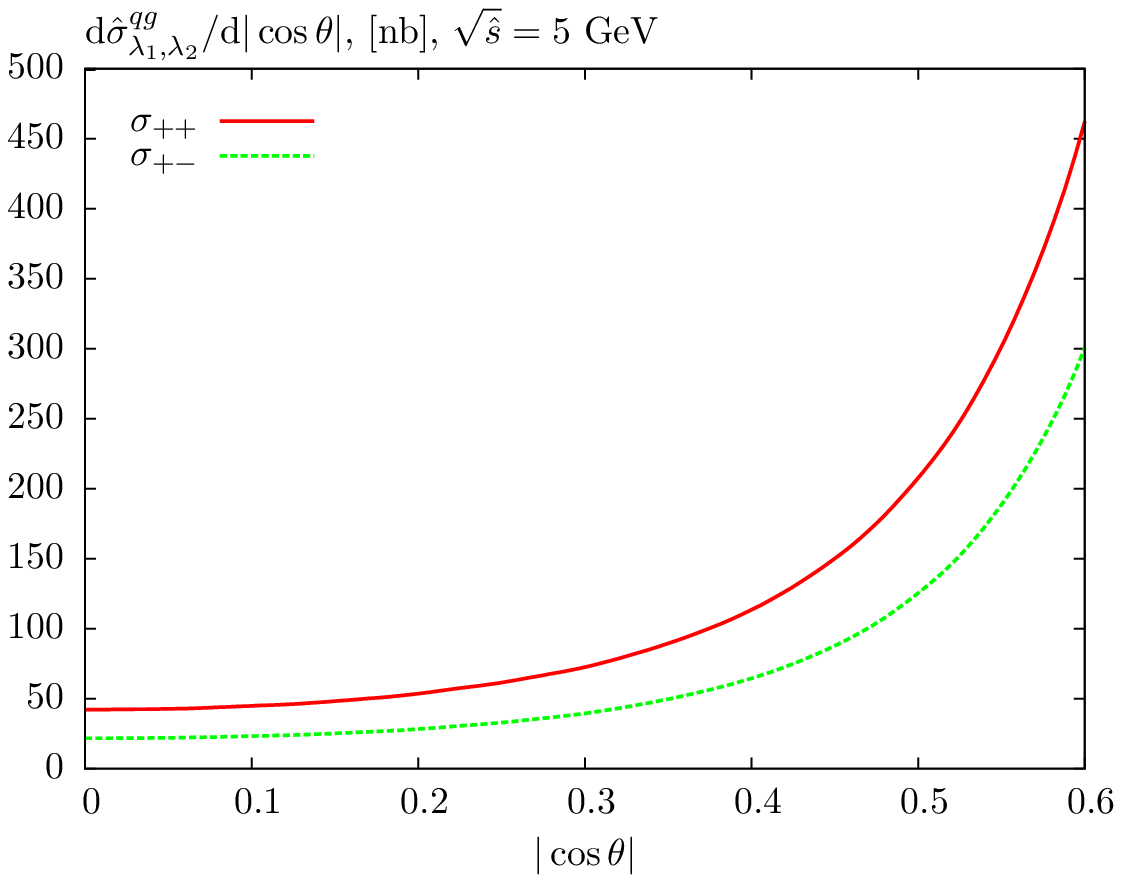}
\includegraphics[scale=0.6]{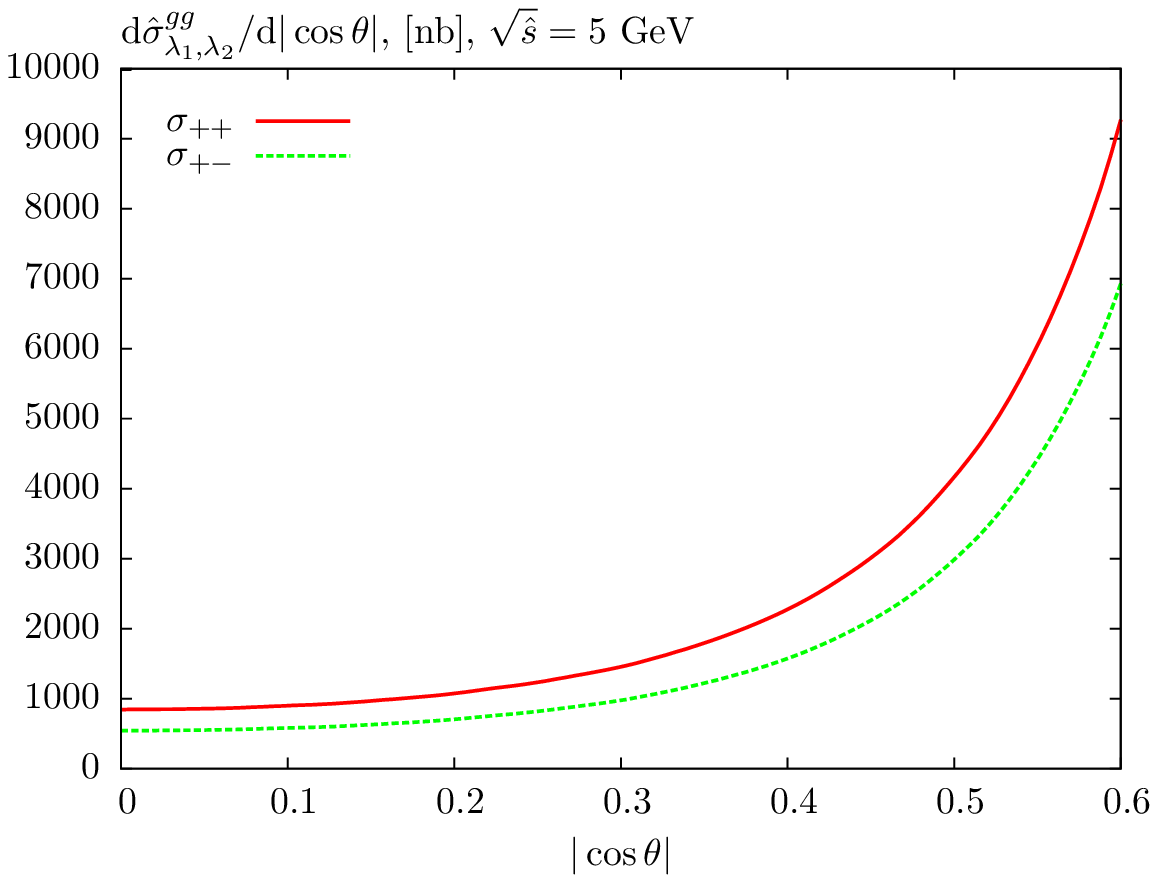}
\caption{Differential cross section ${\rm d}\sigma_{\lambda_1\lambda_2}/{\rm d}|\cos\theta|$ for $gg\to\eta_1 \eta_1$ via $gg \to ggq\overline{q}$ and $gg \to gggg$ diagrams, shown for illustration, for $\sqrt{\hat{s}}=5$ GeV. The $q\overline{q}$ distribution amplitude is given by (\ref{CZ}), with $n_f=3$, while the $gg$ distribution amplitude is given by (\ref{phigex}), with $f_1=f_\pi$. $\lambda_{1,2}$ are the incoming gluon helicities.}\label{ampplots}
\end{center}
\end{figure}

In the case of the $gg \to ggq\overline{q}$ and $gg \to gggg$ amplitudes for $|J_z|=2$ incoming gluons we can find no simple closed form. However, by numerical evaluation we can see that they exhibit a similar angular behaviour to the $J_z=0$ counterparts. We show this in Fig.~\ref{ampplots}, where we plot the differential cross sections ${\rm d}\sigma_{\lambda_1\lambda_2}/{\rm d}|\cos\theta|$ corresponding to the amplitudes $T^{gq}_{+-}$ and $T^{gg}_{+-}$ (or equivalently, $T^{gq}_{-+}$ and $T^{gg}_{-+}$).

We recall that, due to the selection rule which operates for CEP~\cite{Kaidalov:2003fw} the contribution of the $|J_z|=2$ amplitudes is strongly suppressed. As observed in~\cite{HarlandLang:2011qd}, this is particularly important in the case of flavour--non--singlet meson production. These mesons (which only have a $q\overline{q}$ component), cannot be produced in the `ladder--type' $gg \to q\overline{q}q\overline{q}$ or the $gg \to ggq\overline{q}$ subprocesses discussed above, where the $q\overline{q}$ pairs forming the mesons come from the same quark line. The contributing diagrams are instead of the type shown in Fig.~\ref{feyn1} (left), with the photon pair replaced by gluons, but in this case it was found in~\cite{HarlandLang:2011qd} that the corresponding amplitude vanishes for $J_z=0$ incoming gluons, and so the CEP cross section for flavour--non--singlet meson production is expected to be strongly suppressed.

We note that it can readily be shown from the Feynman rules for fermion fields that we should associate an additional factor of $(-1)$ with the flavour--non--singlet amplitude of Fig.~\ref{feyn1} (a), for the case of incoming gluons, but not with the `ladder' diagrams of Fig.~\ref{feyn2}\footnote{In~\cite{HarlandLang:2011qd}, the opposite assignment was taken. However only the quark contribution was considered, in which only the relative sign between these two types of diagram was important. The quoted cross sections and results are therefore correct.}. That the amplitudes corresponding to these two types of diagrams have a relative minus sign is not in fact surprising: if we consider the case that both $q\overline{q}$ pairs are of the same flavour, the diagram of the type shown in Fig.~\ref{feyn1} (a) is completely equivalent (replacing the photon pair with gluons) to that of Fig.~\ref{feyn2} (a) under the interchange of the identical fermion quarks (or anti--quarks).

Finally, it can be shown that the remarkable fact that (\ref{tgq0}), (\ref{tgg0}) and (\ref{lad0}) are identical up to overall colour and normalization factors is not accidental, but can in fact be understood in a `MHV' framework. We refer the reader to Appendix~\ref{mhvcalc} for references and a detailed discussion of this.

\section{Results}\label{res}

From Section~\ref{explicit} we can see that to first approximation (ignoring the flavour non--singlet component and the $|J_z|=2$ amplitudes), the effect of including a $gg$ component to the $\eta'\eta'$ CEP amplitude will simply be to multiply it by an overall normalization factor $N_{gg}$, given by (omitting the $\mu_F^2$ dependence for simplicity)
\begin{equation}\label{ngg}
 N_{gg}=\frac{\int\,{\rm d}x\,{\rm d}y\, \left(\phi_{1,\eta'}(x) \phi_{1,\eta'}(y) T^{qq}(x,y)+2\phi_{1,\eta'}(x) \phi_G(y) T^{qg}(x,y)+\phi_G(x) \phi_G(y) T^{gg}(x,y)\right)}{\int\,{\rm d}x\,{\rm d}y\, \phi_{1,\eta'}(x) \phi_{1,\eta'}(y) T^{qq}(x,y)}\;,
\end{equation}
with all dependence on the final--state kinematic variables (i.e. $\cos(\theta)$ and $\hat{s}$) dropping out, due to the identity of the corresponding amplitudes, up to overall colour and normalization factors. For example for the $qq$ term we may take
\begin{equation}\label{exa}
\int\,{\rm d}x\,{\rm d}y\, \phi_{1,\eta'}(x) \phi_{1,\eta'}(y) T^{qq}(x,y) \to \frac{1}{N_C}n_f\int\,{\rm d}x\,{\rm d}y\, \frac{\phi_{1,\eta'}(x)\phi_{1,\eta'}(y)}{xy(1-x)(1-y)}\;, 
\end{equation}
where the factor of $n_f$ ($=3$) in the numerator comes from summing over the three quark flavours contributing to each $q\overline{q}$ state, divided by the wavefunction normalizations (\ref{qfock}).

To give an initial estimate of the size of this effect, we recall from (\ref{waves}) that the gluon distribution amplitude can be written in the form 
\begin{equation}\label{geig}
\phi_{G,\eta'}(x,\mu_F^2)= \frac{f_1^{\eta'}}{2\sqrt{N_C}}\sqrt{\frac{C_F}{2n_f}} x(1-x) \sum_{n=2,4\cdots} a^G_{n}(\mu_F^2) C^{5/2}_{n-1}(2x-1)\;.
\end{equation}
Neglecting higher order terms in $n$, we therefore have explicitly
\begin{equation}\label{phigex}
\phi_{G,\eta'}(x,\mu_F^2)\approx\frac{5 f_1^{\eta'}}{3\sqrt{6}}a_2^G(\mu_F^2) x(1-x)(2x-1) \;.
\end{equation}
For the quark distribution, we will consider both the asymptotic (\ref{asym}) and CZ (\ref{CZ}) distribution amplitudes, i.e. taking $a_2^{1}(\mu_0^2)=0$ and $a_2^{1}(\mu_0^2)=2/3$ in (\ref{waves}), respectively, with all higher $n$ terms being zero. We have
\begin{align}
 \int\,{\rm d}x\ \frac{\phi_{1,\eta'}^{\rm asym.}(x)}{x(1-x)}&=\sqrt{3} f_1^{\eta'}\;,\\
\int\,{\rm d}x\ \frac{\phi_{1,\eta'}^{\rm CZ}(x)}{x(1-x)}&=\frac{5}{\sqrt{3}}f_1^{\eta'}\;,\\
\int\,{\rm d}x\ \frac{\phi_{G,\eta'}(x)(2x-1)}{x(1-x)}&=\frac{5 a_2^G}{9\sqrt{6}}f_1^{\eta'}\;.\\
\end{align}
Using these we then have, as in (\ref{exa})
\begin{align}\nonumber
 \int\,{\rm d}x\,{\rm d}y\, \phi_{1,\eta'}^{\rm asym.}(x) \phi_{1,\eta'}^{\rm asym.}(y) T^{qq}(x,y) &\to 3(f_1^{\eta'})^2\;,\\ \nonumber
2\int\,{\rm d}x\,{\rm d}y\, \phi_{1,\eta'}^{\rm asym.}(x) \phi_{G,\eta'}(y) T^{qg}(x,y) &\to \frac{5a_2^G}{3}(f_1^{\eta'})^2\;,\\ \nonumber
 \int\,{\rm d}x\,{\rm d}y\, \phi_{1,\eta'}^{\rm CZ}(x) \phi_{1,\eta'}^{\rm CZ}(y) T^{qq}(x,y) &\to \frac{25}{3} (f_1^{\eta'})^2\;,\\ \nonumber
2\int\,{\rm d}x\,{\rm d}y\, \phi_{1,\eta'}^{\rm CZ}(x) \phi_{G,\eta'}(y) T^{qg}(x,y) &\to \frac{25a_2^G}{9}(f_1^{\eta'})^2\;,\\ \label{ngs}
\int\,{\rm d}x\,{\rm d}y\, \phi_{G,\eta'}(x) \phi_{G,\eta'}(y) T^{gg}(x,y) &\to \frac{25 (a_2^G)^2}{108} (f_1^{\eta'})^2\;,
\end{align}
where again the $\mu_F^2$ dependence is omitted for simplicity, and it is understood that $\phi_{1,\eta'}^{\rm CZ}(x)$ is evaluated at the scale $\mu_0$, so that it is given by (\ref{CZ}).

\begin{figure}
\begin{center}
\includegraphics[scale=0.6]{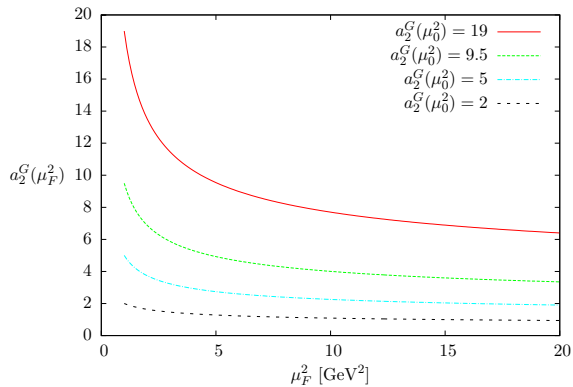}
\caption{$n=2$ coefficient of the expansion (\ref{geig}) of the $gg$ distribution amplitude $\phi_{G,M}(x,\mu_F^2)$, as a function of the scale $\mu_F^2$. We assume that the flavour--singlet $q\overline{q}$ distribution amplitude with which this mixes under evolution is given by the CZ form (\ref{CZ}). The $\mu_F^2$ dependence exhibited in the plot is largely insensitive to the form of the $q\overline{q}$ distribution amplitude. For $a_2^G<0$ the $\mu_F^2$ dependence is identical up to the overall factor of $(-1)$.}\label{b2gplot}
\end{center}
\end{figure}

To give an initial numerical estimate we may then for illustration take the value from~\cite{Kroll:2012hs}
\begin{equation}\label{b2g}
a^G_{2,{\rm fit}}(\mu_0^2)=19 \pm 5\;,
\end{equation}
which is extracted from the transition form factors $F_{\eta(')\gamma}(Q^2)$, and where all higher ($n=4,6$...) order terms are neglected. However, this corresponds to the $gg$ distribution amplitude at the scale $\mu_0^2 = 1\,{\rm GeV}^2$. In fact (see also~\cite{Kroll:2012hs}), the leading coefficient $a_2^G(\mu_F^2)$ of $\phi_G(x,\mu_F^2)$ evolves rapidly  away from its value at $\mu_0$, decreasing by roughly a factor of $\sim 3$ for $\mu_F^2\approx 10\, {\rm GeV}^2$, and decreasing at a much slower rate after that, see Fig.~\ref{b2gplot}. Thus, to give an initial numerical estimate we can take $a^G_n(\mu_F^2=10 \,{\rm GeV}^2) \approx 7$, corresponding to an experimentally realistic value of $M_X\approx 5$ GeV. In this case, we find from (\ref{ngs}) that
\begin{align}\label{nggex}
N_{gg}^{\rm asym}&\approx 9\;, \\
N_{gg}^{\rm CZ}&\approx 5\;,
\end{align}
and therefore we will expect a potentially large increase ($\sim N_{gg}^2$) in the $\eta(')\eta(')$ CEP cross section for this value (\ref{b2g}) of $a^G_2(\mu_0^2)$. While (\ref{b2g}) may give some rough guidance for the expected size of the $gg$ component of the $\eta(')$, we note that the fit of~\cite{Kroll:2012hs} contains important uncertainties, see the discussion at the end of this section, and so the question of the size of $a_2^G(\mu_0^2)$ is not settled. Therefore, to give a conservative evaluation of the sensitivity of the CEP process to the size of this $gg$ component, we will consider in what follows a band of cross section predictions, corresponding to the range
\begin{equation}\label{b2g1}
 a_2^G(\mu_0^2)\in (-a^G_{2,{\rm fit}}/2,+a^G_{2,{\rm fit}}/2)=(-9.5,9.5)\;.
\end{equation}
As we will see, even within this quite narrow and conservative range of values, the predicted CEP cross section changes considerably. The results we present below are based on an exact numerical calculation, using a modification of the \texttt{SuperCHIC} MC~\cite{SuperCHIC}, and including the precise effect of the evolution of the $gg$ and $q\overline{q}$ wavefunctions, in contrast to the rough estimate described above. We take $\mu_0=1$ GeV and $\mu_F^2=M_X^2/2$ (being of order the squared momentum transfer $\sim |\hat{t}|, |\hat{u}|$) throughout, although other choices of factorization scale are of course possible.

%

\begin{figure}
\begin{center}
\includegraphics[scale=0.6]{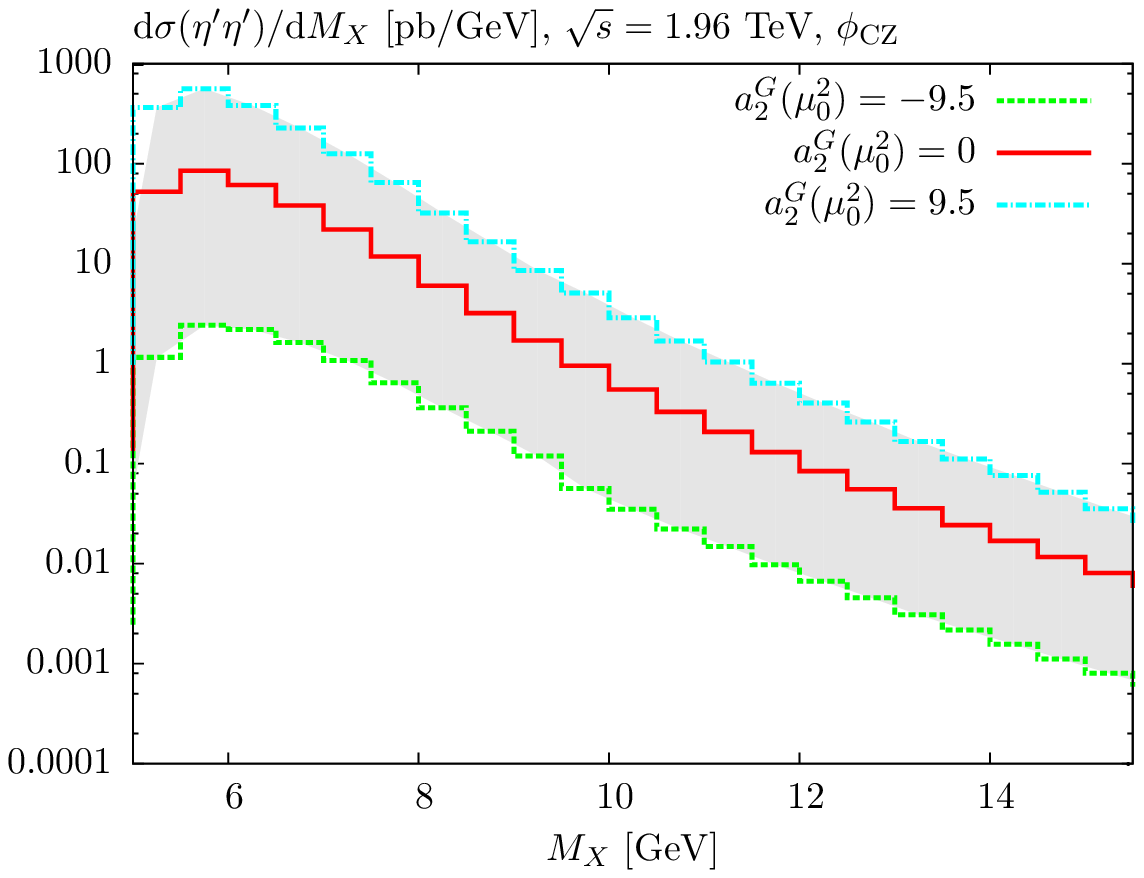}
\includegraphics[scale=0.6]{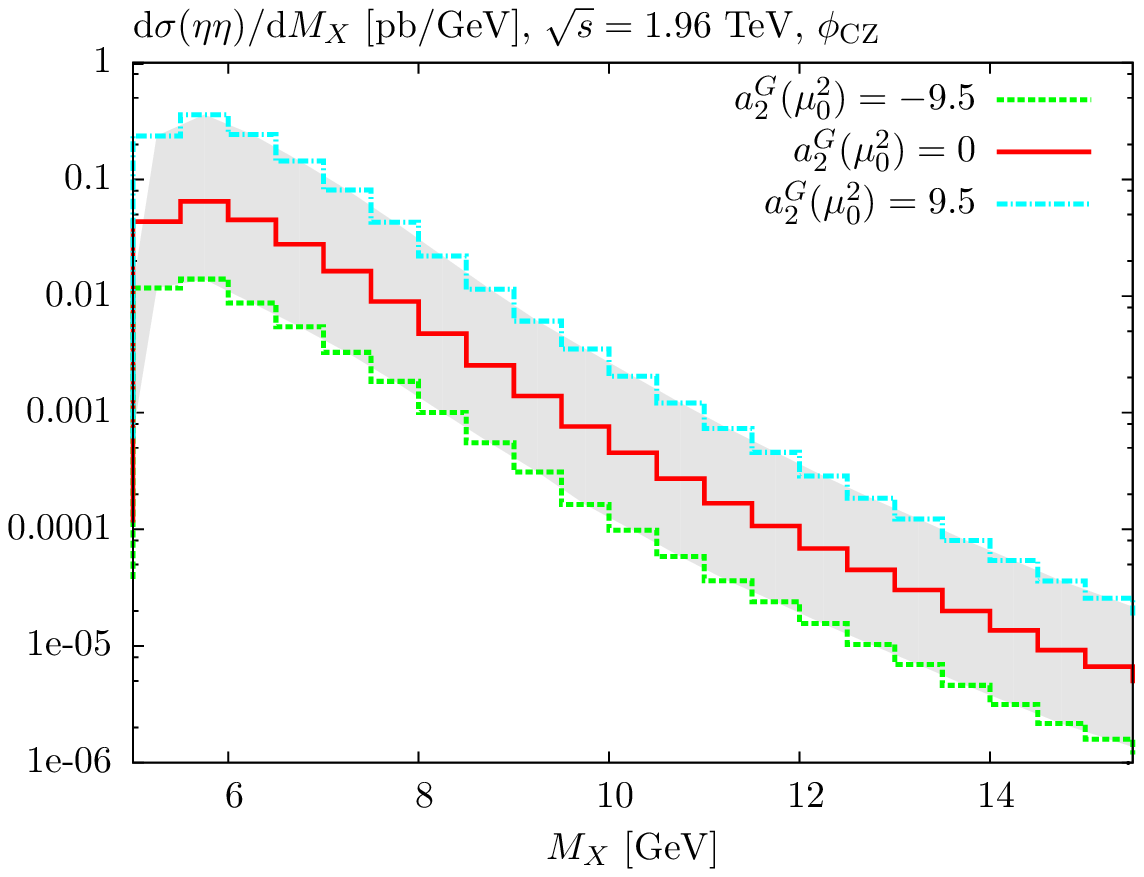}
\includegraphics[scale=0.6]{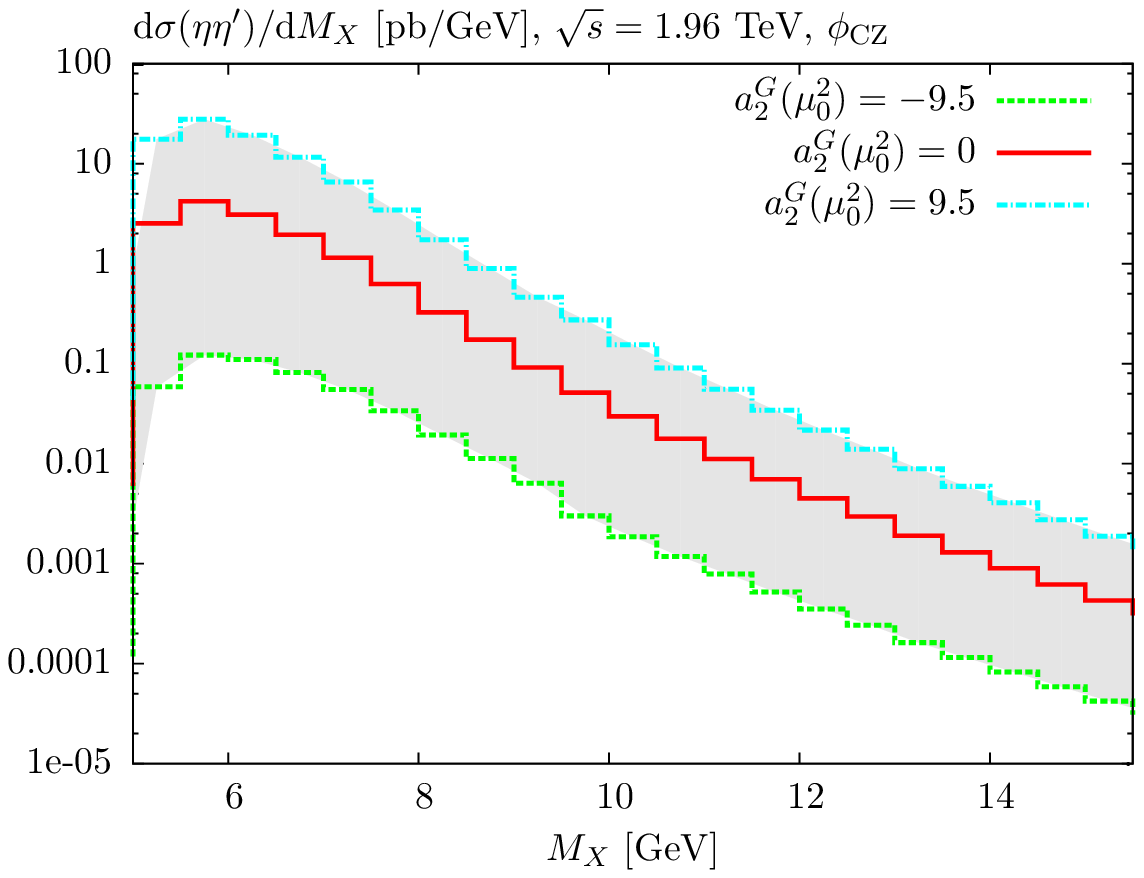}
\caption{Differential cross section ${\rm d}\sigma/{\rm d}M_X$ for $X=\eta'\eta',\eta\eta,\eta\eta'$ production  at $\sqrt{s}=1.96$ TeV with MSTW08LO PDFs~\cite{Martin:2009iq}, taking the CZ form (\ref{CZ}) for the quark distribution amplitude, and for a band of $a_2^G(\mu_0^2)$ values for the $gg$ distribution amplitude. The mesons are required to have transverse energy $E_\perp>2.5$ GeV and pseudorapidity $|\eta|<1$.}\label{etamcz}
\end{center}
\end{figure}

\begin{figure}
\begin{center}
\includegraphics[scale=0.6]{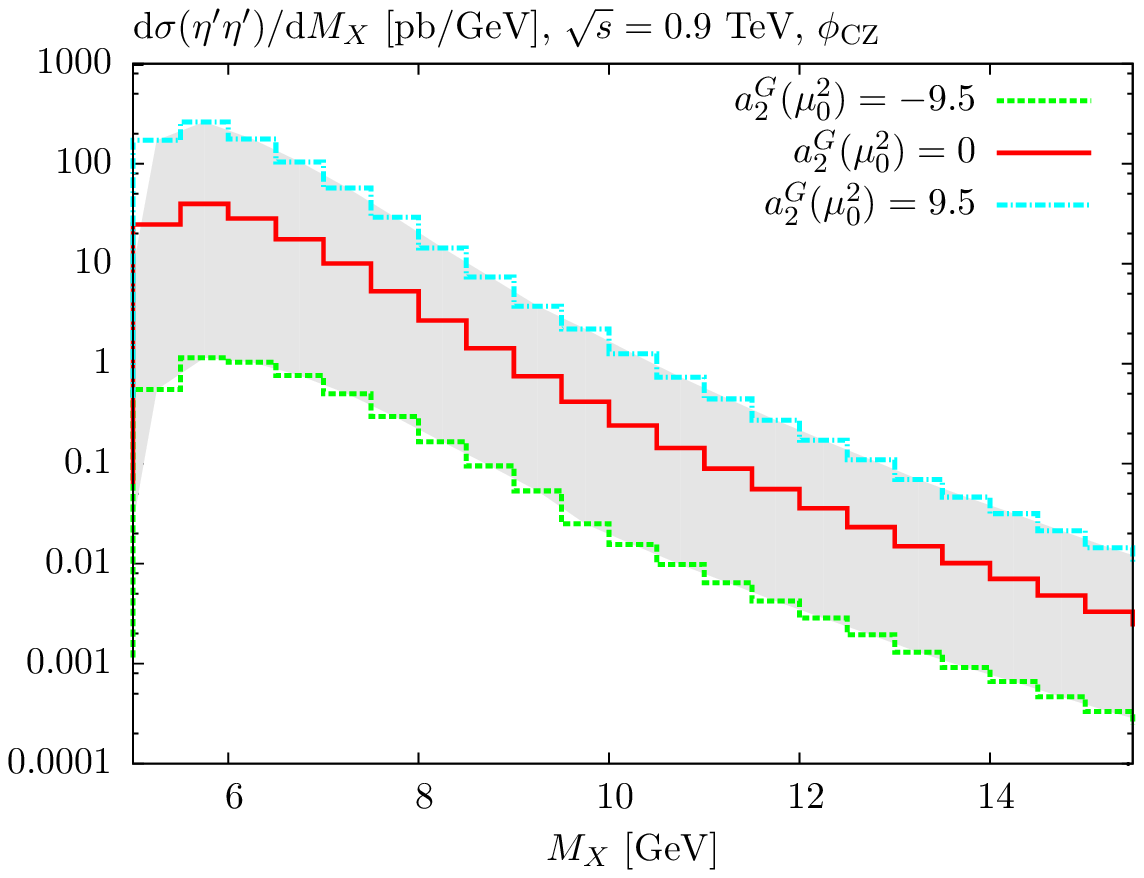}
\includegraphics[scale=0.6]{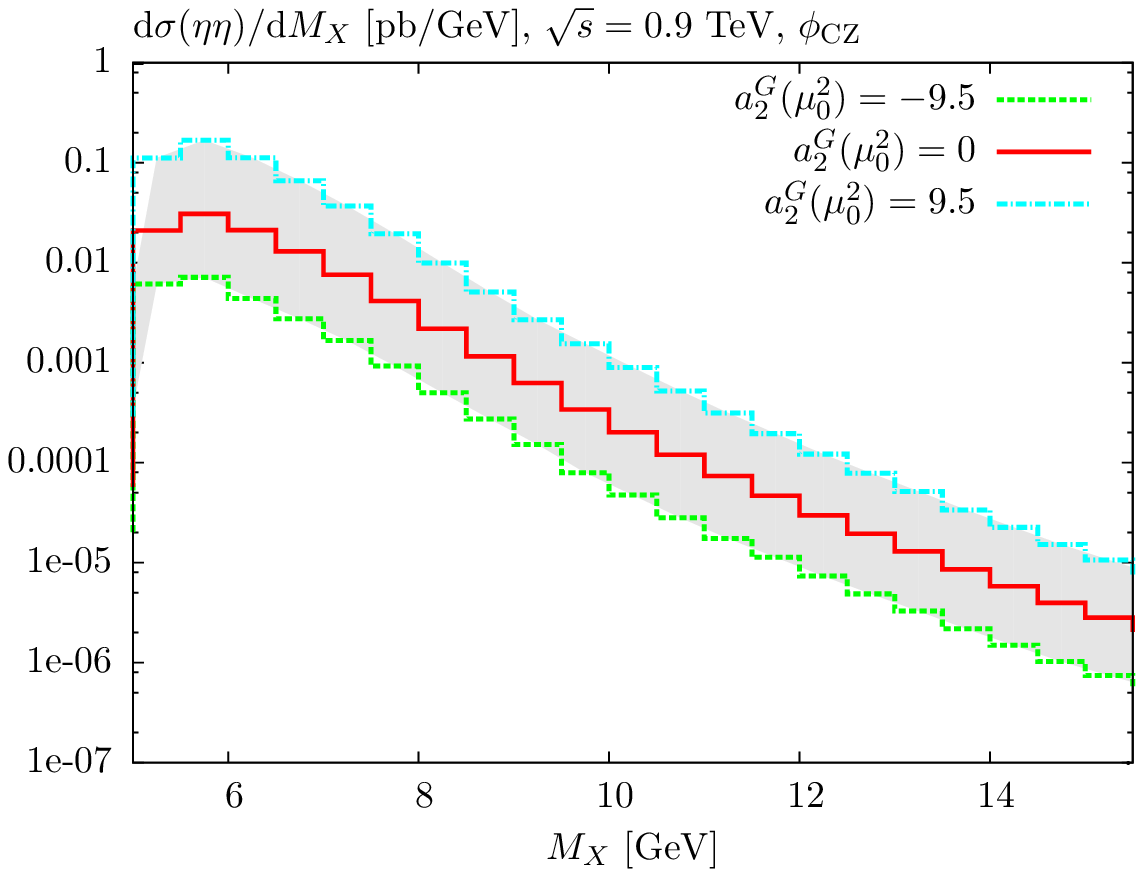}
\includegraphics[scale=0.6]{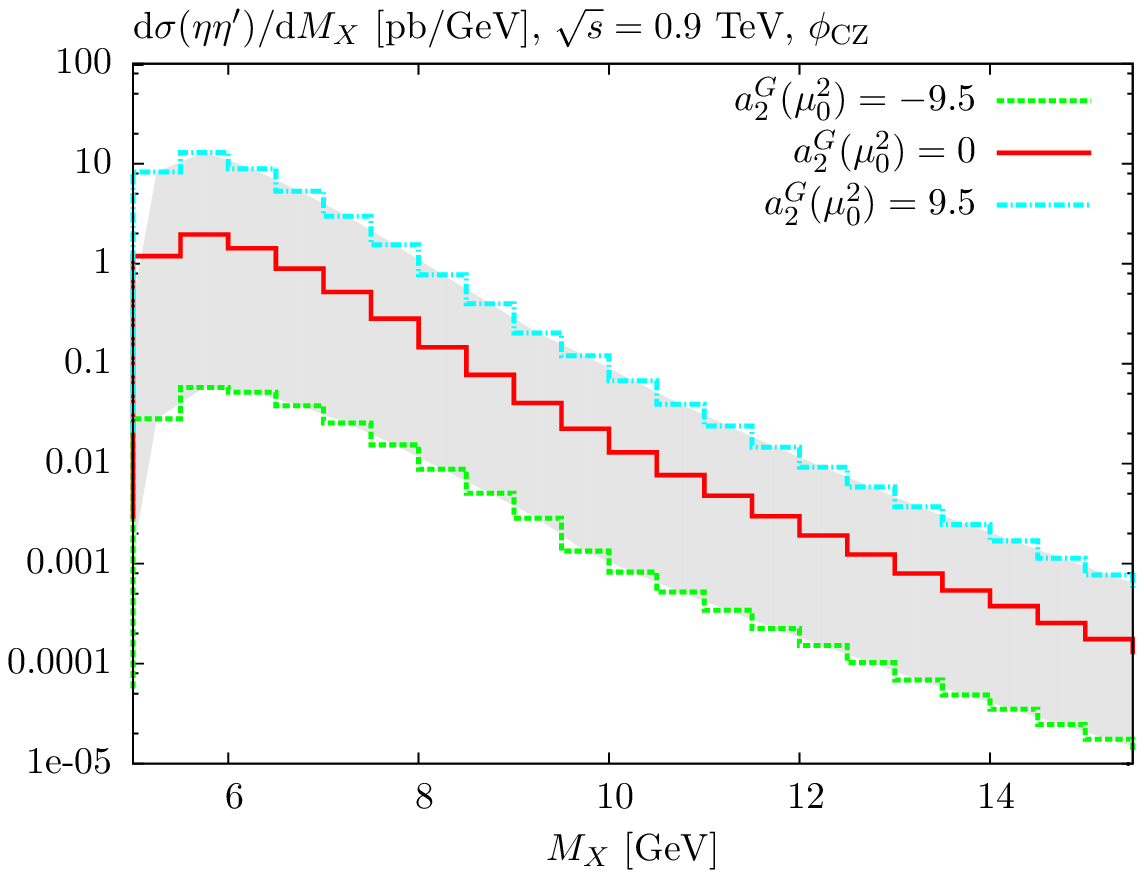}
\caption{Differential cross section ${\rm d}\sigma/{\rm d}M_X$ for $X=\eta'\eta',\eta\eta,\eta\eta'$ production  at $\sqrt{s}=0.9$ TeV with MSTW08LO PDFs~\cite{Martin:2009iq}, taking the CZ form (\ref{CZ}) for the quark distribution amplitude, and for a band of $a_2^G(\mu_0^2)$ values for the $gg$ distribution amplitude. The mesons are required to have transverse energy $E_\perp>2.5$ GeV and pseudorapidity $|\eta|<1$.}\label{etamcz900}
\end{center}
\end{figure}

We show in Figs.~\ref{etamcz},~\ref{etamcz900} the cross sections for $\eta'\eta'$, $\eta\eta$ and $\eta\eta'$ production using the CZ form (\ref{CZ}) for the quark distribution amplitude, for three different choices $a_2^G(\mu_0^2)=(-9.5,0,9.5)$ of the $gg$ distribution amplitude normalization (\ref{b2g1}), and for the mesons required to have transverse energy $E_\perp>2.5$ GeV and pseudorapidity $|\eta|<1$. All cross section predictions in this paper are calculated using MSTW08LO PDFs~\cite{Martin:2009iq}: while there is in general a reasonably large variation in the predicted cross sections (which are sensitive to the low $x$ and $Q^2$ region of the gluon density) on the choice of PDF, we use this set as it gives a prediction which is consistent with the existing CDF exclusive $\gamma\gamma$ data~\cite{Aaltonen:2011hi}, see for example~\cite{HarlandLang:2012qz} for further discussion. We show the cross sections for $\sqrt{s}=0.9, 1.96$ GeV, as these are the energies at which the existing CDF data on exclusive final--states have been taken, and from a further analysis of which we may hope for an observation of these meson pair CEP processes to come~\cite{albrow}. At the LHC, we expect the cross section (for the same event selection) to be roughly a factor of $\sim 3$--5 larger for $\sqrt{s}=7$--14 TeV, with the particle distributions almost unchanged. For $a_2^G(\mu_0^2)=9.5$, we can see that the expected cross section increases significantly, while for negative $a_2^G$ the $gq$ interferes destructively with the $gg$ and $qq$ contributions, leading to a suppression in the predicted cross section. The predicted $\eta\eta$ cross section is less suppressed, due to the $|J_z|=2$ flavour--non--singlet contribution (see the discussion in section~\ref{explicit}), which for the lower flavour--singlet cross sections (corresponding to $a_2^G=0,-9.5$) becomes important. 

\begin{figure}
\begin{center}
\includegraphics[scale=0.6]{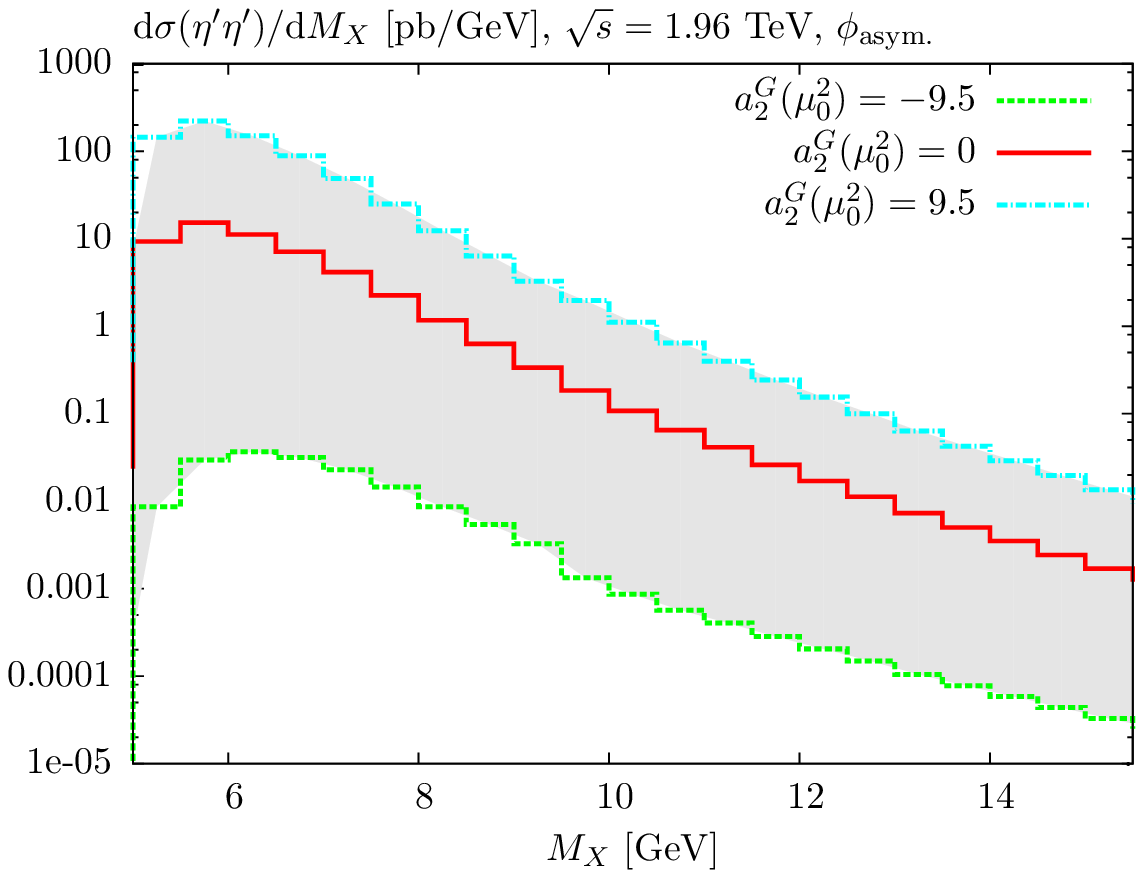}
\includegraphics[scale=0.6]{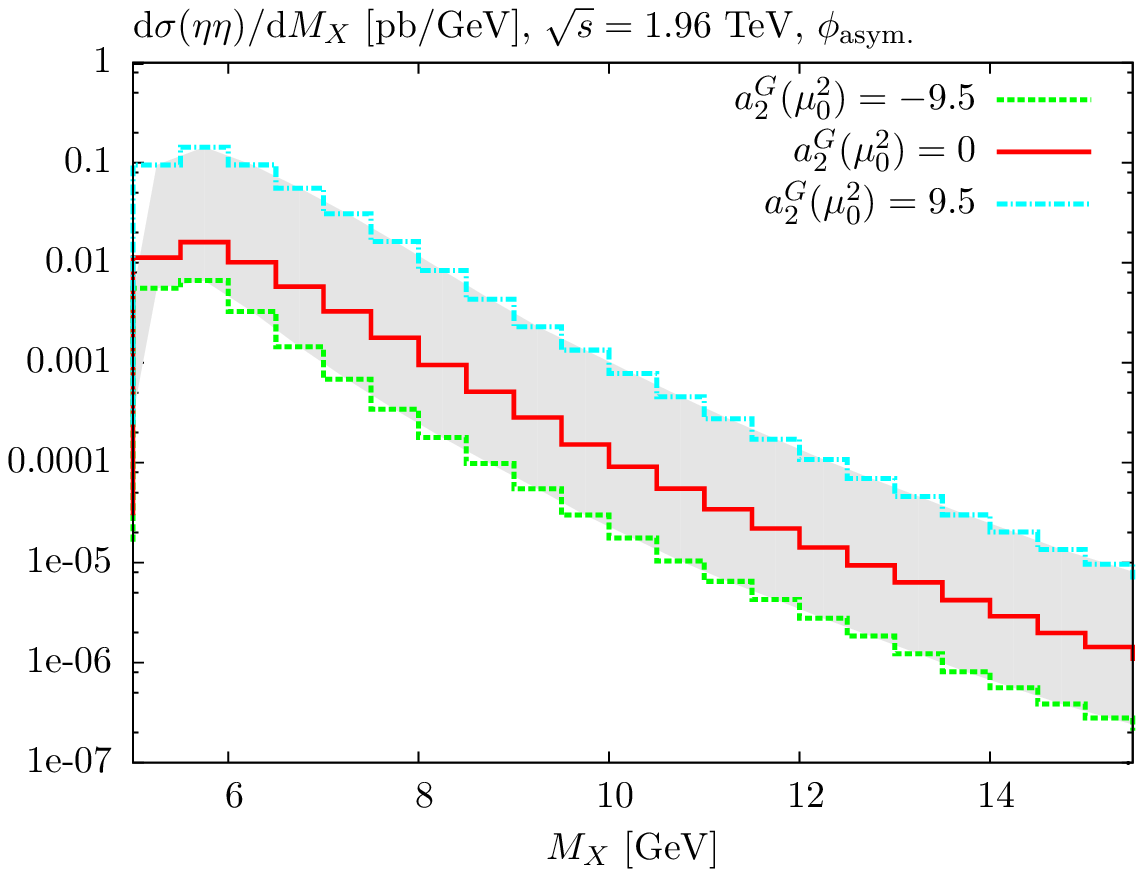}
\includegraphics[scale=0.6]{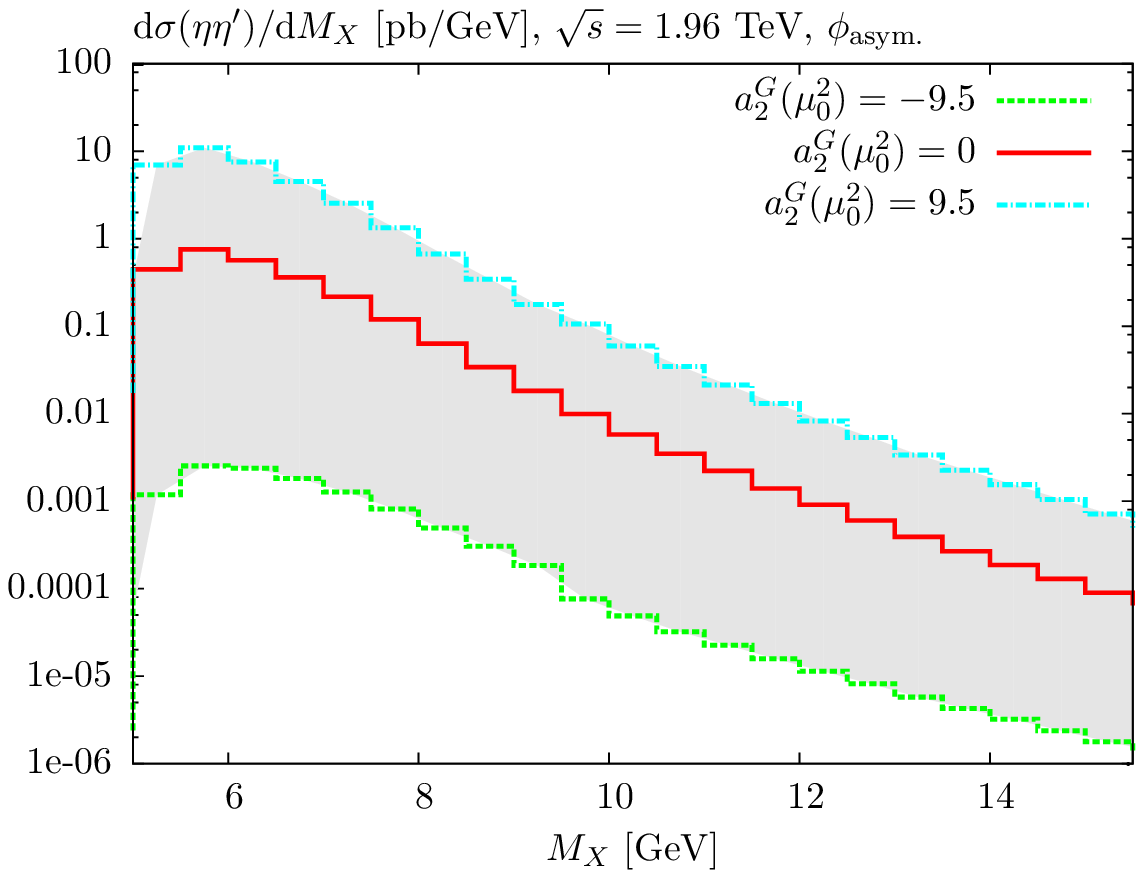}
\caption{Differential cross section ${\rm d}\sigma/{\rm d}M_X$ for $X=\eta'\eta',\eta\eta,\eta\eta'$ production  at $\sqrt{s}=1.96$ TeV with MSTW08LO PDFs~\cite{Martin:2009iq}, taking the asymptotic form (\ref{asym}) for the quark distribution amplitude, and for a band of $a_2^G(\mu_0^2)$ values for the $gg$ distribution amplitude. The mesons are required to have transverse energy $E_\perp>2.5$ GeV and pseudorapidity $|\eta|<1$.}\label{etamas}
\end{center}
\end{figure}

We note that the extracted value of $a_2^G(\mu_0^2)$ in general depends on the form of the flavour--singlet quark distribution amplitude, i.e. the size of the $a_n^1(\mu_0^2)$ in (\ref{waves}). With this in mind, in Fig.~\ref{etamas} we show the same cross sections at $\sqrt{s}=1.96$ TeV, but for the asymptotic form of the quark distribution amplitude (\ref{asym}). The $a_2^G(\mu_0^2)=0$ cross section is predicted to be somewhat smaller than for the CZ distribution amplitude choice, while a similar, somewhat larger, relative enhancement is seen for the case that $a_2^G(\mu_0^2)=9.5$. For $a_2^G(\mu_0^2)=-9.5$ the destructive interference is in fact almost exact for this $M_X$ region, and so the cross sections are predicted to be very strongly affected. However, we would not necessarily expect to see such a large destructive interference effect in the data, as such an exact cancellation may not occur at higher orders. Again, we stress that these only correspond to specific cases in a band of possible $a_2^G(\mu_0^2)$, and that the real size of this may be smaller.

Neglecting contributions with $|J_z|=2$ incoming gluons, the ratio of $\eta'\eta'$ to $\eta\eta$ ($\eta\eta'$) cross sections are determined by the mixing in (\ref{etafit}--\ref{mixing}) to be 
\begin{align}\nonumber
\sigma(\eta'\eta'):\sigma(\eta\eta'):\sigma(\eta\eta)&=1:2\tan^2(\theta_1):\tan^4(\theta_1)\;,\\ \label{ratcross}
&\approx 1:\frac{1}{19}:\frac{1}{1450}\;,
\end{align}
where we have taken the value of $\theta_1$ from (\ref{thetafit})\footnote{We note that the predicted $\eta'\eta$ and $\eta\eta$ cross sections are lower than in~\cite{HarlandLang:2011qd}. This is due to the different choice of mixing scheme (\ref{etafit}--\ref{mixing}) and in particular the lower value of $\theta_1$, which leads to a smaller flavour--singlet component of the $\eta$. It is found in for example~\cite{Feldmann:1998vh,Feldmann:2002kz,Escribano:2005qq} that this scheme (\ref{etafit}--\ref{mixing}) and choice of mixing parameters describe the available data well. A measurement of the cross section ratios in (\ref{ratcross}) would certainly shed  further light on this.} and the factor of `2' in the $\eta\eta'$ case accounts for the non--identity of the final--state particles. A measurement of the cross section ratios (in particular, $\sigma(\eta'\eta')/\sigma(\eta\eta')$), would therefore serve as a probe of the mixing parameter, $\theta_1$, see (\ref{thetafit}). This ratio is predicted to be unchanged by the inclusion of a non--zero gluon component $a_2^G\neq 0$, as for $J_z=0$ incoming gluons only the flavour--singlet component $\eta_1$ of the $\eta$ and $\eta'$ mesons contributes, while we have seen in Section~\ref{fcalcsec} that the $gg \to ggq\overline{q}$ and $gg \to gggg$ amplitudes are identical in form to the purely quark case $gg \to q\overline{q}q\overline{q}$, and will therefore only effect the overall normalization of this flavour--singlet contribution. In the ratio of cross sections, this overall factor cancels and we are left with the scaling of (\ref{ratcross}) irrespective of the size of the $gg$ component. In Table~\ref{etarats} we show numerical results for the ratios (\ref{ratcross}): due to the relative importance of the $|J_z|=2$ flavour non--singlet contribution, in the $\eta\eta$ case, this scaling is only expected to be approximate, see below (there is also in all cases a small effect due to the differing $\eta$ and $\eta'$ masses in the MC simulation).

Thus, to first approximation we can only look at absolute value of the various $\eta(')\eta(')$ CEP cross sections to determine the size of the $gg$ component, $a_2^G(\mu_0^2)$. This is potentially problematic because of the other uncertainties in the CEP calculation, due primarily to the value of the survival factors $S^2_{\rm eik}$, $S^2_{\rm enh}$, which are not known precisely, and potential higher--order corrections in the hard process, which combined are expected to a give a factor of $\sim {}^\times_\div 2-3$ uncertainty, as well as a sizeable PDF uncertainty in the low--$x$, $Q^2$ regime relevant to such processes (see~\cite{HarlandLang:2012qz} for a detailed discussion of this and further references). Nevertheless, given the sensitivity of the CEP cross section to the $gg\to gggg$ and $gg \to ggq\overline{q}$ subprocess, if the $gg$ component of the $\eta(')$ is sizeable enough (as the result of~\cite{Kroll:2012hs}, for example, suggests), such a measurement may still provide useful information. 

However, it is more reliable to look at the ratio of the $\eta(')\eta(')$ cross section to other processes, in which case many of the uncertainties due to PDFs and survival factors largely cancel out and a potentially much cleaner measurement of the $gg$ component of the $\eta(')$ becomes possible. With this in mind we show for illustration in Table~\ref{etarats1} the ratio of the $\eta\eta$ to $\pi^0\pi^0$ and $\eta'\eta'$ to $\gamma\gamma$ cross sections\footnote{The $\pi^+\pi^-$ CEP cross section is predicted to be identical to the neutral pion case, up to a factor of $2$ due to the non--identity of the final--state particles. The observation of such a process has recently been reported in~\cite{MAdif12}, although it is not clear that this measurement probes sufficiently high pion $k_\perp$ for the perturbative approach discussed here to be applicable. In order to better clarify the situation, a measurement of the meson $k_\perp$-- distributions corresponding to the same data would be very useful, while a comparison between the $k_\perp$ (and $M_{\pi\pi}$) distributions at 1.96 TeV and 900 GeV would probe the size of any possible contamination due to proton dissociation~\cite{HarlandLang:2013jf}. We also note that we expect $\pi^+\pi^-$ (and $K^+K^-$) data to come soon from CMS~\cite{enterria}, with a veto applied on any additional particles within their rapidity coverage, which should contain a sizeable exclusive component.}, for the same three choices $a_2^G(\mu_0^2)=(-9.5,0,9.5)$ of the $gg$ distribution amplitude normalization (\ref{b2g1}).

As mentioned above, the scaling of (\ref{ratcross}) is only approximate, as it ignores the possibility of a flavour--non--singlet contribution for the case that the incoming gluons in the $gg \to \eta(')\eta(')$ subprocess are in a $|J_z|=2$ state. Such a contribution may in particular be relevant for $\eta\eta$ production, when the suppression in the flavour--singlet cross section due to the small mixing $\sim\sin^4 \theta_1$ can be comparable to the suppression of the flavour--non--singlet contribution due to the $J_z=0$ selection rule that operates for CEP~\cite{Kaidalov:2003fw}. This can be seen in Table~\ref{etarats}, where the ratio of $\eta'\eta'$ to $\eta\eta$ (and $\eta\eta'$) cross sections differs from (\ref{ratcross}) due to this additional contribution. For larger values of $a_2^G(\mu_0^2)$, the flavour non--singlet contribution becomes relatively less important and we approach the expected values; a measurement of these ratios may therefore serve as an additional probe 
of the $gg$ contribution. Recalling that the $\pi^0\pi^0$ CEP cross section is also predicted to be strongly suppressed, due to the vanishing at LO of the $gg \to \pi^0\pi^0$ amplitude for $J_z=0$ incoming gluons, a measurement of the ratios $\sigma(\eta(')\eta('))/\sigma(\pi^0\pi^0)$ would also represent as an important probe of the $J_z=0$ selection rule\footnote{It should be noted that any NNLO corrections or higher twist effects which allow a $J_z=0$ contribution may cause the precise value of the flavour--non--singlet $\eta\eta$ (and $\pi^0\pi^0$) cross section to be somewhat larger than the leading--order, leading--twist $|J_z|=2$ estimate.}.

We can also see from Table~\ref{etarats1} (see also~\cite{HarlandLang:2011qd}) that the $\eta'\eta'$ cross section is expected to be somewhat larger than for $\gamma\gamma$ CEP. However, multiplying by the corresponding branching ratios, ${\rm Br} (\eta' \to \gamma\gamma)\approx 2\%$ and ${\rm Br} (\eta \to \gamma\gamma)\approx 40\%$~\cite{Beringer:1900zz}, squared, we can see in Table~\ref{etarats1} that the cross sections for $\eta'\eta'$ (and also for $\eta\eta$) CEP after branching to the $4\gamma$ final state are predicted to be a small fraction of the direct $\gamma\gamma$ CEP cross section for the relevant event selection, with a similar result holding for the $\eta\eta'$ final state. We therefore do not expect these to represent an important background to the existing CDF\cite{Aaltonen:2011hi} and any forthcoming CMS~\cite{CMSgam} $\gamma\gamma$ data\footnote{Moreover there should be a  further reduction, in any contamination from $\eta,\eta'\to 4\gamma$ due to the experimental exclusivity and isolation cuts and event selection. On the other hand it is worth mentioning that the multi--photon decay modes of the $\eta$ and $\eta'$ are quite sizeable (about 72\% and 18\% respectively~\cite{Beringer:1900zz}) and, in principle, these may also contribute to the background.}.

\begin{table}[h]
\begin{center}
\begin{tabular}{|l|c|c|c|}
\hline
$a_2^G(\mu_0^2)$&-9.5&0&9.5\\
\hline
$\sigma(\eta'\eta')/\sigma(\eta\eta)$&210&1300&1600\\
\hline
$\sigma(\eta'\eta')/\sigma(\eta\eta')$&20&20&20\\
\hline
$\sigma(\eta\eta')/\sigma(\eta\eta)$&11&66&78\\
\hline
\end{tabular}
\caption{Ratios of $\eta(')\eta(')$ CEP cross sections at $\sqrt{s}=1.96$ TeV with MSTW08LO PDFs~\cite{Martin:2009iq}, for a $gg$ distribution amplitude with different choices of $a_2^G(\mu_0^2)$ and with the $q\overline{q}$ distribution amplitude given by the CZ form (\ref{CZ}). The meson are required to have transverse energy $E_\perp>2.5$ GeV and pseudorapidity $|\eta|<1$.}\label{etarats}
\end{center}
\end{table}

\begin{table}[h]
\begin{center}
\begin{tabular}{|l|c|c|c|}
\hline
$a_2^G(\mu_0^2)$&-9.5&0&9.5\\
\hline
$\sigma(\eta\eta)/\sigma(\pi^0\pi^0)$&2.7&12&66\\
\hline
$\sigma(\eta'\eta')/\sigma(\pi^0\pi^0)$&570&16000&100000\\
\hline
$\sigma(\eta'\eta')/\sigma(\gamma\gamma)$&3.5&100&660\\
\hline
$\sigma(\eta'\eta'\to 4 \gamma)/\sigma(\gamma\gamma)$&0.0017&0.049&0.33\\
\hline
$\sigma(\eta\eta \to 4 \gamma)/\sigma(\gamma\gamma)$&0.0025&0.012&0.066\\
\hline
\end{tabular}
\caption{Ratios of $\eta(')\eta(')$ to $\pi^0\pi^0$ and $\gamma\gamma$ CEP cross sections at $\sqrt{s}=1.96$ TeV with MSTW08LO PDFs~\cite{Martin:2009iq}, for a $gg$ distribution amplitude with different choices of $a_2^G(\mu_0^2)$ and with the $q\overline{q}$ distribution amplitude given by the CZ form (\ref{CZ}). The meson/photons are required to have transverse energy $E_\perp>2.5$ GeV and pseudorapidity $|\eta|<1$. Also show are the ratios $\sigma(\eta'\eta')/\sigma(\gamma\gamma)$ and $\sigma(\eta\eta)/\sigma(\gamma\gamma)$ cross sections multiplied by the $\eta(') \to \gamma\gamma$ branching ratios squared.}\label{etarats1}
\end{center}
\end{table}

We recall that while various estimates are available in the literature (see for example~\cite{Thomas:2007uy,Ke:2011fj,Mathieu:2009sg,Ambrosino:2009sc,Escribano:2007cd,DiDonato:2011kr}), no firm consensus exists about the precise size of the $gg$ component\footnote{For example, the situation with regards to the $\chi_{c(0,2)}$ decays into $\eta$ and $\eta'$ pairs appears somewhat puzzling. Experimentally, no enhancement in the production of the $\eta,\eta'$ pairs as compared to pions is observed (after taking trivial phase space effects into account). This may indicate that there is some destructive interference between the $q\overline{q}$ and the $gg$ components of the pseudoscalar bosons, or that the $gg$ component is small. See~\cite{Kroll:2012hs} for other ideas and~\cite{Ochs:2013gi} for a review and further discussion.} of the $\eta(')$, which is indicative of the uncertainty present in these `standard' approaches, see for example~\cite{Thomas:2007uy} for a discussion of the theoretical uncertainties (
in e.g. decay form factors) present in such extractions. In~\cite{Kroll:2012hs}, for example, the gluonic contribution to the $\eta(')$ transition form factor $F_{\eta(')\gamma}(Q^2)$ only enters at NLO, and so has a relatively small effect, so that the extracted value relies on a precision fit to the data\footnote{The potential importance of power corrections to $F_{\eta(')\gamma}(Q^2)$ at lower values of $Q^2$, where much of the existing data lies, may in fact cast doubt on the reliability of such a fit. We thank Viktor Chernyak for a useful discussion on this topic.}. In the case of CEP, on the other hand, the gluonic amplitudes (\ref{tgq0}), (\ref{tgg0}) enter at LO and so, as we have seen above, it is potentially strongly sensitive to such a $gg$ contribution. An observation of $\eta \eta$, $\eta'\eta'$ and/or $\eta\eta'$ CEP could therefore shed important light on this uncertain issue. 

\section{Conclusion and Outlook}\label{conc}

In this paper we have performed a detailed analysis of the central exclusive pair production of pseudoscalar $\eta$ and $\eta'$ mesons, $pp\to p\,+\,X\,+\,p$, with $X=\eta\eta$, $\eta'\eta'$ and $\eta\eta'$. We have concentrated on the case that the mesons are produced at sufficiently high transverse momentum $k_\perp$, that a perturbative approach which combines the Durham model of the general CEP process (see e.g.~\cite{HarlandLang:2010ep} and references therein) with the `hard exclusive' formalism (see e.g.~\cite{Brodsky:1981rp,Benayoun:1989ng,Chernyak:2006dk} and references therein) used to model the specific $gg\to \eta(')\eta(')$ subprocess, can be applied.

We have in particular extended the previous work of~\cite{HarlandLang:2011qd} to include the possibility of a non--zero flavour--singlet $gg$ valence component of the $\eta'$ (and $\eta$) mesons. We recall that knowledge of the quark and gluon components of these states would provide important information about various aspects of non--perturbative QCD, see the Introduction for more details. With this in mind we have calculated for the first time the amplitudes $gg \to gg q\overline{q}$ and $gg \to gggg$, where the $gg$, $q\overline{q}$ pairs form collinear pseudoscalar meson states with the right spin and colour quantum numbers, see Fig.~\ref{feyn2}.

It is worth emphasizing that the parton--level amplitudes relevant to the CEP of flavour--singlet (and non--singlet) meson pairs in the perturbative regime exhibit some remarkable features. In~\cite{HarlandLang:2011qd} it was shown that, for the case that  the incoming gluons are in a $J_z=0$ state, the parton--level amplitude for the production of flavour--non--singlet states, $gg \to q\overline{q} q\overline{q}$, vanishes completely at LO, while the $|J_z|=2$ amplitude exhibit a `radiation zero' at a particular value of the c.m.s. scattering angle. On the other hand, the production of flavour--singlet $q\overline{q}$ states can take place via a separate set of `ladder diagrams', see Fig.~\ref{feyn2} (a), and in this case the $J_z=0$ amplitude does not vanish, and can be written in a very simple form. In this paper, we have shown that if one or both of these $q\overline{q}$ pairs forming the flavour--singlet meson states is replaced by a gluon pair $gg$, as in e.g. Fig.~\ref{feyn2} (b,c), then these 
apparently unrelated amplitudes, $gg \to gg q\overline{q}$ and $gg \to gggg$, are identical to each other and to this `ladder--type' amplitude, up to overall normalization factors. That is, they are predicted to have exactly the same angular dependence in the incoming $gg$ rest--frame. In~\cite{HarlandLang:2011qd} and in Section~\ref{mhvcalc} we have shown how these remarkable and non--trivial results may be explained in the MHV framework by the fact that the same external parton orderings contribute in all three cases.

In CEP, this contribution from a $gg$ valence component of the $\eta'$, $\eta$ mesons enters at the same (leading) order to the purely $q\overline{q}$ contribution. We have seen in Section~\ref{res} that as a result the predicted $\eta(')\eta(')$ CEP cross sections display quite a strong sensitivity to any potential $gg$ component of the $\eta'$, $\eta$ mesons. An observation of $\eta\eta$, $\eta'\eta'$ and/or $\eta\eta'$ CEP could therefore provide a potentially powerful handle on the structure of the $\eta'$, $\eta$ states. More generally, the observation of such meson pair ($\eta(')\eta(')$, $\pi\pi$, $KK$...) CEP processes would serve as a test of the unique properties of the $gg\to M\overline{M}$ parton--level helicity amplitudes, in particular the vanishing of the $J_z=0$ amplitudes for flavour non--singlet mesons and the identical form of the $gg$ and $q\overline{q}$ mediated flavour--singlet amplitudes. We emphasize in particular that the $\eta'\eta'$ CEP cross section is predicted to be 
very large, see Section~\ref{res}. These results, in conjunction with the $J_z=0$ CEP selection rule, lead to highly non--trivial predictions for the meson pair CEP cross sections, which it would be very interesting to compare to data.

Finally, we note that we may expect new results on $\eta(')\eta(')$ and meson pair CEP more generally to come from further analysis of the existing CDF data (in particular the  existing $4\gamma$ candidates with $E_{T}> 2.5$ GeV and forward rapidity gaps) as well as from the CMS/Totem (ATLAS) special low-pileup  runs with sufficient luminosity~\cite{albrow,MAdif12}. At CDF, the observation of $\gamma\gamma$ CEP has already been reported~\cite{Aaltonen:2011hi}, and it was determined experimentally that the contamination caused by $\pi^0\pi^0\to 4\gamma$   CEP, with the photons in the $\pi^0$ decay merging or one photon being undetected, is very small ($< 15$ events, corresponding to a ratio $N(\pi^0\pi^0)/N(\gamma\gamma)<0.35$, at 95\% C.L.), in agreement with the prediction of~\cite{HarlandLang:2011qd}, for which $\sigma(\pi^0\pi^0)/\sigma(\gamma\gamma)\sim 1\%$. This prediction is a non--trivial result of the perturbative CEP framework and the hard exclusive formalism, and an observation of $\pi^0\pi^0$ CEP, which may hopefully come with the increased statistics that a further analysis of the existing data can bring~\cite{albrow}, would certainly represent an interesting further test of the theoretical formalism. In Section~\ref{res} we observed that the $\eta'\eta'$ and $\eta\eta$ to $4\gamma$ cross sections are expected to be of about the same size or maybe even be larger than the $\pi^0\pi^0$ cross section, but that they should not be an important background to $\gamma\gamma$ CEP. Nonetheless, this raises the possibility of a future observation of these processes via the $\gamma\gamma$ decay chain. We also note that the branching ratios for the  $\rho^0\gamma$ and $\eta \pi^+\pi^-$ decays of the $\eta'$ are sizeable, and may be viable channels for the observation of $\eta'\eta'$ (and $\eta\eta'$) CEP, a possibility which would hopefully be confirmed after further analysis and simulations.

To conclude, the CEP of $\eta'$ and $\eta$ meson pairs, in the perturbative regime, represents a novel (and complementary) probe of the size of a flavour--singlet $gg$ component of these mesons. It is our hope that future $\eta'$, $\eta$ pair CEP data and analysis will be forthcoming from the Tevatron and the LHC and that through this we can shed some light on this interesting and currently uncertain question.

\section*{Acknowledgements}
We thank Mike Albrow, Erik Brucken, Victor Chernyak, Risto Orava, Kornelija Passek--Kumeri\u{c}ki and Antoni Szczurek for useful discussions. This work was supported by the grant RFBR 11-02-00120-a and by the Federal Program of the Russian State RSGSS-4801.2012.2. WJS is grateful to the IPPP for an Associateship. VAK thanks the Galileo Galilei Institute for Theoretical Physics for hospitality and the INFN for partial support during the completion of this work.

\appendix
\section{Meson distribution amplitudes: anomalous dimensions}\label{ap1}

The anomalous dimensions which control the evolution (\ref{b2gev}) are given by~\cite{Ohrndorf:1981uz,Baier:1981pm}
\begin{align}\nonumber
\gamma_n^{qq}&=C_F\left[3+\frac{2}{(n+1)(n+2)}-4\sum_{i=1}^{n+1}\frac{1}{i}\right]\;,\\ \nonumber
\gamma_n^{qg}&=C_F\frac{n(n+3)}{3(n+1)(n+2)}\qquad n\geq 2\;,\\ \nonumber
\gamma_n^{gq}&=n_f\frac{12}{(n+1)(n+2)}\qquad n\geq 2\;,\\
\gamma_n^{gg}&=\beta_0+N_C\left[\frac{8}{(n+1)(n+2)}-4\sum_{i=1}^{n+1}\frac{1}{i}\right]\;.
\end{align}
In terms of these we can then write down the eigenvalues $\gamma_{n}^{(\pm)}$ which diagonalise the anomalous dimensions
\begin{equation}
 \gamma_{n}^{(\pm)}=\frac{1}{2}\left[\gamma_n^{qq}+\gamma_n^{gg}+\sqrt{(\gamma_n^{qq}-\gamma_n^{gg})^2+4\gamma_n^{qg}\gamma_n^{gq}}\right]\;.
\end{equation}
The parameters $\rho_n^{(\pm)}$ in (\ref{b2gev}) are given by
\begin{equation}
 \rho_n^{(+)}=6\frac{\gamma_n^{gq}}{\gamma_n^{(+)}-\gamma_n^{gg}}\;,\qquad  \rho_n^{(-)}=\frac{1}{6}\frac{\gamma_n^{qg}}{\gamma_n^{(-)}-\gamma_n^{qq}}\;,
\end{equation}

\section{MHV calculation}\label{mhvcalc}

It is well known (see for example~\cite{Mangano:1990by}) that the tree level $n$--gluon scattering amplitudes, in which the maximal number ($n-2$) of gluons have the same helicity, the so--called `maximally helicity violating' (MHV), or `Parke--Taylor', amplitudes, are given by remarkably simple formulae~\cite{Parke:1986gb,Berends:1987me}. These results were extended using supersymmetric Ward identities to include amplitudes with one and two quark--antiquark pairs in~\cite{Mangano:1990by}, where `MHV' refers to the case where ($n-2$) partons have the same helicity. In these cases, simple analytic expressions can again be written down for the MHV amplitudes, while for greater than 2 fermion--anti--fermion pairs (recalling that the helicities of a connected fermion--anti--fermion pair must be opposite) no MHV amplitudes exist. 

With this in mind, we will show that the fact that (\ref{tgq0}), (\ref{tgg0}) and (\ref{lad0}) are identical up to overall colour and normalization factors is not accidental, but follows from the observation that the $J_z=0$ helicity amplitudes considered in Section~\ref{explicit} are MHV, with $n-2=4$ partons (the two incoming gluons, and two outgoing partons) having the same helicity. We will in particular show that the same set of MHV partial amplitudes, i.e. with the same orderings of the parton momenta, contributes in all cases. This section explores purely theoretical aspects of the previously calculated amplitudes, and so a reader who is only interested in the CEP cross section predictions may skip forward to Section~\ref{res}.

We recall that in general it is well known that a full $n$--parton amplitude $\mathcal{M}_n$ can be written in the form of a `dual expansion', as a sum of products of colour factors $T_n$ and purely kinematic partial amplitudes $A_n$
\begin{equation}\label{mhv}
\mathcal{M}_n(\{p_i,h_i,c_i\})=ig^{n-2}\sum_\sigma T_n(\sigma\{c_i\})A_n(\sigma\{1^{\lambda_1},\cdots,n^{\lambda_n})\;,
\end{equation}
where $c_i$ are colour labels, $i^{\lambda_i}$ corresponds to the $i$th particle ($i=1\cdots n$), with momentum $p_i$ and helicity $\lambda_i$, and the sum is over appropriate simultaneous non--cyclic permutations $\sigma$ of colour labels and kinematics variables. The purely kinematic part of the amplitude $A_n$ encodes all the non--trivial information about the full amplitude, $\mathcal{M}_n$, while the factors $T_n$ are given by known colour traces, see for instance~\cite{Georgiou:2004wu} for more details. 
Adjusting the notation from (\ref{mhv}) slightly for clarity, the $n$--gluon and $q\overline{q}$ ($(n-2)$--gluon) MHV partial amplitudes are given by
\begin{align}\label{mhvpartg}
A(g_1^+,g_2^+,...,g_i^-,...,g_j^-,...,g_n^+)&=\frac{\langle i\, j \rangle^4}{\prod_{k=1}^n \langle k\, k+1 \rangle}\;,\\ \label{mhvpartq}
A(g_1^+,g_2^+,...,g_i^-,...,\overline{q}_j^-,q_{j+1}^+,...,g_n^+)&=\frac{\langle i\, j \rangle^3\langle i\, j+1 \rangle}{\prod_{k=1}^n \langle k\, k+1 \rangle}\;,
\end{align}
where $\langle k_i\,k_j\rangle \equiv \langle k_i^-|k_j^+\rangle = \overline{u}_-(k_i)u_+(k_j)=\overline{v}_+(k_i)v_-(k_j)$ is the standard spinor contraction, and all momenta are defined as incoming. For the case that the quark (anti--quark) has positive (negative) helicity, it is sufficient to simply interchange $j$ and $j+1$ in the numerator of the right hand side. The form of (\ref{mhvpartq})  expresses the important requirement that gluons are always emitted from the `same side' of the connected quark--antiquark line (see for example Fig.~1 of ~\cite{Wu:2004jxa} and the discussion in the text); that is, the $q\overline{q}$ pair must appear consecutively and in the same order. 

We can then apply these general formulae to the specific $n=6$ amplitudes in Section~\ref{explicit}, where the outgoing $q\overline{q}$ and $gg$ pairs must form collinear meson states of the correct colour and spin. Firstly, we can see that in the case of both of the $n$--parton amplitudes (\ref{mhvpartg}) and (\ref{mhvpartq}), the numerators are independent of the particular (non--cyclic) permutation of the partons being considered, and will therefore factorize. The form this takes is given by the corresponding spin projections, as in (\ref{cpp}) and (\ref{cgg}). In particular, if we define the following momenta
\begin{equation}\label{mom}
l_3=xp_3\qquad l_4=(1-x)p_3 \qquad l_5=yp_4 \qquad l_6=(1-y)p_4\;,
\end{equation}
where $p_{3,4}$ are the 4--momenta of the outgoing mesons, then the numerator in the 6--gluon case corresponds to (recalling that the helicity here is defined with respect to the incoming gluon momenta)
\begin{align}\nonumber
&(g^-(l_3)g^+(l_4)-g^+(l_3)g^-(l_4))(g^-(l_5)g^+(l_6)-g^+(l_5)g^-(l_6))\\ \nonumber
&\to \langle 4 \,6\rangle^4+\langle 3\, 5\rangle^4-\langle 3\, 6\rangle^4-\langle 4 \,5\rangle^4\\ \label{num}
&=\hat{s}^2(2y-1)(2x-1)\;.
\end{align}
In the $q\overline{q}$ case, the final expression is the same, but with the factor of `$(2y-1)$' removed.

We can see from (\ref{mhvpartg}) and (\ref{mhvpartq}) that the amplitudes for a given ordering of partons are identical between the two cases, up to these overall numerator factors. To justify the statement that it is indeed the same set of orderings which contribute in the case of both amplitudes (\ref{tgq0}) and (\ref{tgg0}), we must consider the colour factors, $T_n$, in (\ref{mhv}). In the $q\overline{q}$ case this is given by
\begin{equation}\label{tnq}
T_n((n-2)g+q\overline{q})=(\lambda^{1} \cdots \lambda^{n-2})_{i_1 j_1}\;,
\end{equation}
where $i_1(j_1)$ is the colour index the quark (anti--quark). In the $n$--gluon case, $T_n$ is given by
\begin{equation}
T_n(ng)={\rm Tr}\,(\lambda^1\cdots \lambda^n)\;.
\end{equation}
For a given $T_n$, the corresponding diagram has the same cyclic ordering of the quark and gluons as their colour labels in $T_n$. The $\lambda$ matrices are normalized (unconventionally) so that ${\rm Tr}(\lambda^a\lambda^b)=\delta^{ab}$, as required by the form of the expansion given in (\ref{mhv}).

\begin{figure}
\begin{center}
\subfigure[]{\includegraphics[scale=0.8]{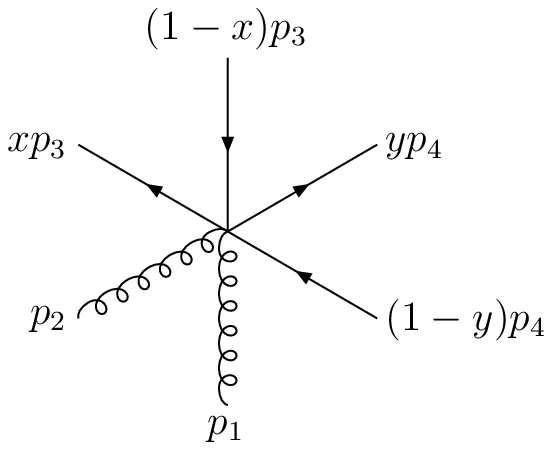}}
\subfigure[]{\includegraphics[scale=0.8]{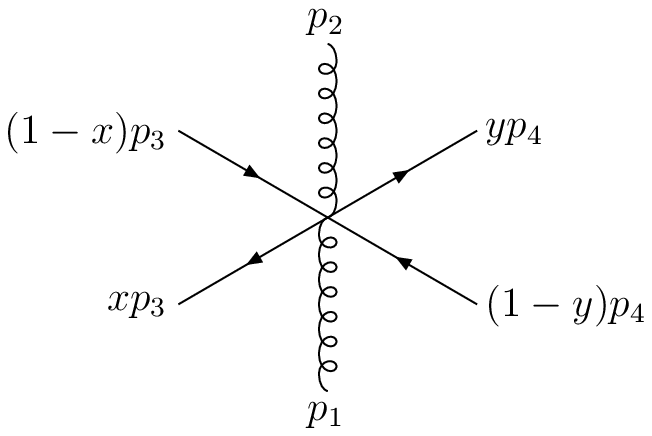}}
\subfigure[]{\includegraphics[scale=0.8]{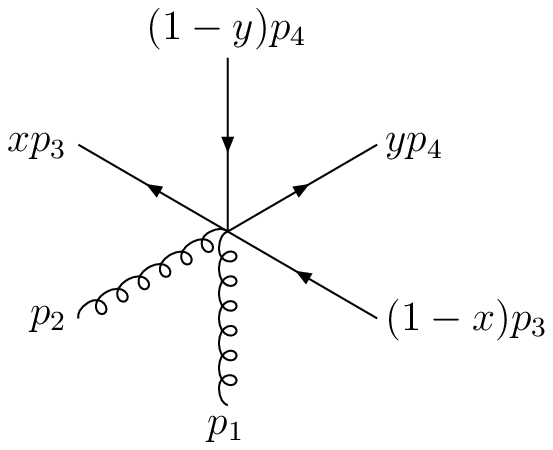}}
\subfigure[]{\includegraphics[scale=0.8]{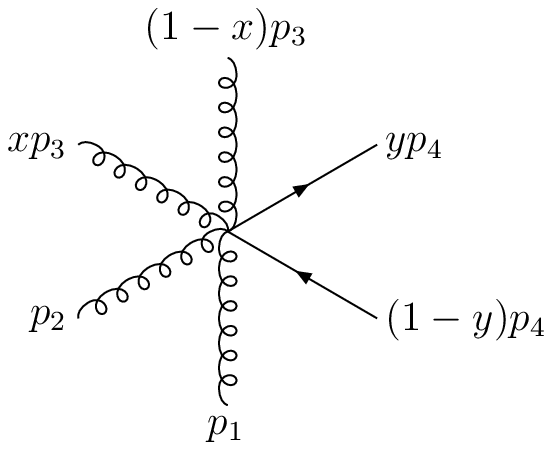}}
\subfigure[]{\includegraphics[scale=0.8]{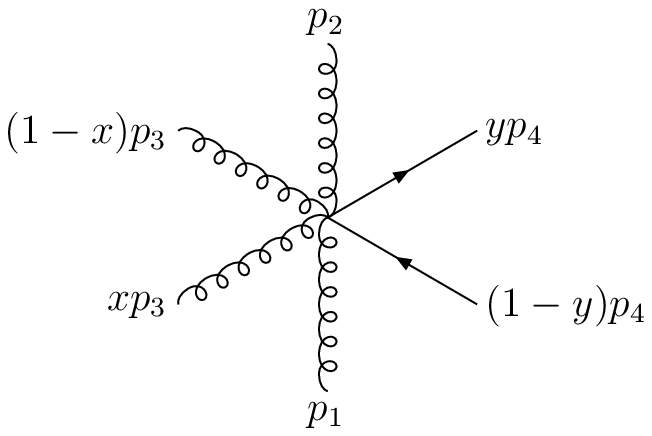}}
\subfigure[]{\includegraphics[scale=0.8]{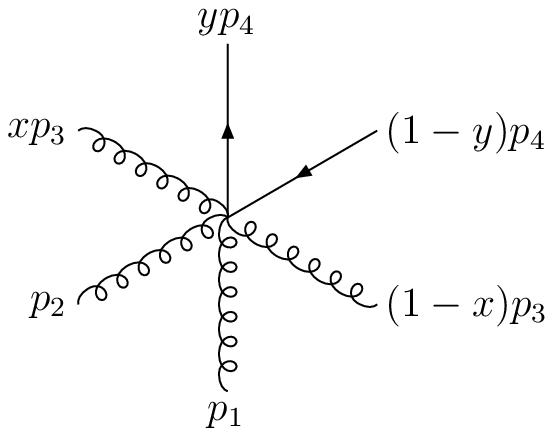}}
\subfigure[]{\includegraphics[scale=0.8]{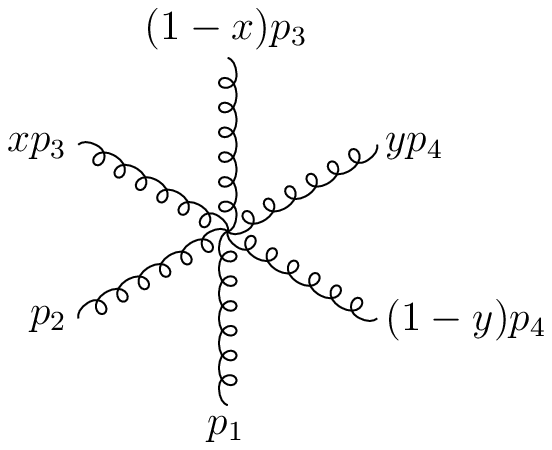}}
\subfigure[]{\includegraphics[scale=0.8]{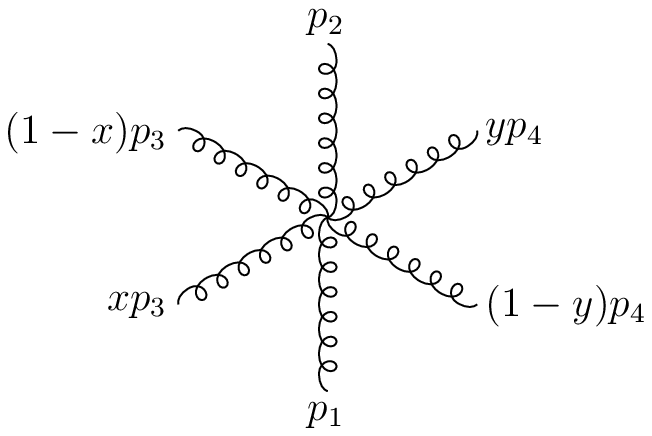}}
\subfigure[]{\includegraphics[scale=0.8]{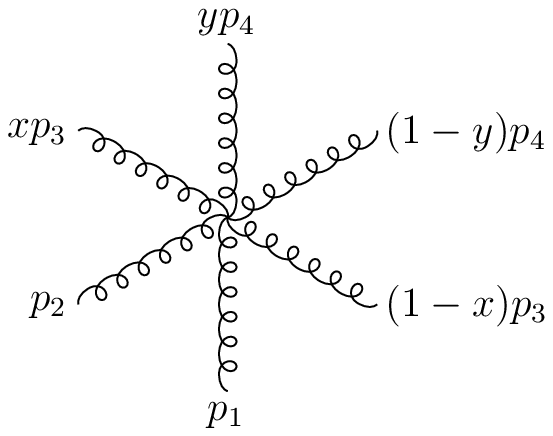}}
\caption{Representative contributing MHV diagrams for flavour singlet meson pair production, via (a--c) $gg \to q\overline{q}q\overline{q}$ (d--f) $gg \to q\overline{q}gg$ (g--i) $gg \to gggg$}\label{mhvall}
\end{center}
\end{figure}

Now, an explicit calculation in Section~\ref{explicit} has shown that in both cases it is only the leading terms in $N_C$ which give a non--zero contribution. In the $gg \to q\overline{q}gg$ case the terms of order $1/N_C$ which come from certain `abelian diagrams' (i.e. those which are identical to the $\gamma\gamma \to q\overline{q}gg$ diagrams, with the photons replaced by gluons) do not contribute due to the vanishing of the $J_z=0$ two--photon amplitudes (\ref{tggam0}), while in the $gg \to 4g$ case, all colour factors coming from the colour--singlet final and initial states are of order $N_C^2$. We therefore know that all the partial amplitudes with colour factors $T_n$ which are sub--leading in $N_C$ must sum to give zero contribution, and can be neglected. By considering the various possible values of $T_n$ for different non--cyclic orderings of the quarks and gluons, it is easy to show that only those diagrams of the type shown in Fig.~\ref{mhvall} (d--f) and Fig.~\ref{mhvall} (g--i) give such a leading $N_C$ colour factor, with in particular
\begin{align}
T_6(4g+q\overline{q})&=N_C+\cdots\;,\\
T_6(6g)&=N_C^2+\cdots\;,
\end{align}
in all cases. That is, the same type of diagrams (corresponding to a given ordering of the external parton momenta) contribute in both cases. Recalling that the denominator term in (\ref{mhvpartg}) and (\ref{mhvpartq}) is identical for a given particle ordering, while the numerator and colour factors factorize, we can see that the resulting amplitudes will be equivalent up to these overall factors, as expected.

Explicitly, calculating the partial amplitudes corresponding to the $6g$ diagrams shown in Fig.~\ref{mhvall} (g--i), including those coming from interchanging $p_1 \leftrightarrow p_2$, $x \to 1-x$ and $y \to 1-y$, we find
\begin{align}\label{ta}
A_g^{6g}&=-2\frac{(\hat{t}-\hat{u})^2}{\hat{s}\hat{u}^2\hat{t}^2}\;,\\ \label{tb}
A_h^{6g}&=-2\frac{\hat{s}}{\hat{u}^2\hat{t}^2}\;,\\ \label{tc}
A_i^{6g}&=0\; ,
\end{align}
where these correspond to the kinematic amplitudes with the numerator factors, as in (\ref{num}), as well as an additional factor of $xy(1-x)(1-y)$ in the denominator, omitted for simplicity, and the subscript (`g,h,i') indicates the corresponding diagram in Fig.~\ref{mhvall}. The amplitudes in the $4g$ case are all a factor of $2$ smaller due to the requirement that only those diagrams with the gluons emitted on one side of the quark line are included, while there is no such requirement for the purely gluonic case. 


In fact, the individual MHV amplitudes corresponding to these parton orderings are divergent, due to factors of $\langle l_3l_4\rangle\sim p_3^2=0$ and $\langle l_5 l_6\rangle\sim p_4^2=0$ present in the denominators, but the sum over all particle interchanges (in particular, of the two gluons forming the meson states, i.e. $l_3(l_5) \leftrightarrow l_4(l_6)$, where there is only one pair to be interchanged in the $ggq\overline{q}$ case) is not. To arrive at these (finite) results requires a careful grouping of the contributing amplitudes and application of the Schouten identity
\begin{equation}\label{sc}
\langle 1\,2\rangle \langle 3\,4\rangle =\langle 1\,4\rangle \langle 3\,2\rangle +\langle 1\,3\rangle \langle 2\,4\rangle\;.
\end{equation}
Combining (\ref{tc}) with the relevant numerator, colour and normalization factors, we can readily reproduce the expressions of (\ref{tgg0}) and (\ref{tgq0}), up to an overall phase, which we take to be due the differing convention for the overall phase of the amplitude which the normalization of (\ref{mhv}) implies.

Finally we turn to the $gg \to q\overline{q}q\overline{q}$ amplitude. We first consider the colour factor, $T_n$, in (\ref{mhv}) for the $q\overline{q}q\overline{q}$ case. This is given by
\begin{equation}\label{tnq2}
T_n((n-4)g+2q\overline{q})=\frac{(-1)^p}{N_C^p}(\lambda^{a_1} \cdots \lambda^{a_l})_{i_1 \alpha_1}(\lambda^{b_1} \cdots \lambda^{b_{l'}})_{i_2 \alpha_2}\;,
\end{equation}
where $i_1,i_2$ are the colour indices of the quarks, $\alpha_1,\alpha_2$ are the colour indices of the anti--quarks and the labels $a_i$, $b_i$ refer to the gluons. The pair $(\alpha_1 \alpha_2)$ is a permutation of the pair $(j_1 j_2)$, where quark $i_k$ is connected by a fermion line to antiquark $j_k$. $p$ is then the number of times $\alpha_k=j_k$, with $p=1$ if $(\alpha_1 \alpha_2)\equiv(j_1 j_2)$. The sum is over all the partitions of the gluons ($l+l'=n-4$, $l=0,\cdots,n-4$) and over the permutations of the gluon indices, with the product of zero $\lambda$ matrices given by a Kronecker delta. 

In the case of (\ref{lad0}), we can again see that it is only the leading terms in $N_C$ ($O(1)$), which contribute to the hard amplitude. Again, it can readily be shown that the relevant partial amplitudes for this have the same parton ordering as those in the $6g$ and $4g$ cases, as shown in Fig.~\ref{mhvall} (a--c). Diagrams a and b correspond to the $p=0$ and c to $p=1$ in (\ref{tnq2}), with the additional factor of $1/N_C$ in the case of diagram c canceled by the factor of $\delta_{ii}$ in the numerator; in both cases, the colour factor is then $O(1)$. In the case of diagram c both gluons must be emitted by one of the quark lines, otherwise the amplitude will vanish due a factor of ${\rm Tr}(\lambda)$ in the numerator. 

The kinematic amplitude for two non--identical quark anti--quark pairs, $q\overline{q}$ and $Q\overline{Q}$, is given by (see e.g.~\cite{Birthwright:2005vi})
\begin{equation}\label{a2g}
 A(q_1^{h_1},g_2^+,\cdots,\overline{Q}_m^{-h_2},Q_{m+1}^{h_2},g_{m+2}^+,\cdots,\overline{q}_n^{-h_1})=\frac{F(h_1,h_2)\langle 1\, m \rangle \langle n\, m+1 \rangle}{\prod_{k=1}^n \langle k\, k+1 \rangle}\;,
\end{equation}
where
\begin{align}\nonumber
F(+,+)&=\langle m\, n \rangle^2  & F(+,-) =\langle n\, m+1 \rangle^2 \;,\\ \label{f2g}
F(-,+)&=\langle 1\, m \rangle^2  & F(-,-) =\langle 1\, m+1\rangle^2\;.
\end{align}
We note that (\ref{a2g}) and (\ref{f2g}) are for the case that $p=0$ in (\ref{tnq2}), i.e. for diagrams a and b of Fig.~\ref{mhvall}. For the ($p=1$) case of diagram c, we must simply interchange $n \leftrightarrow m$ in the above expressions. We can again readily show that summing over the contributing quark helicity states according to the spin projection (\ref{cpp}), as was done in (\ref{num}) for the $6g$ case, will simply give an overall factor of $\hat{s}^2$ for the numerator of (\ref{a2g}), irrespective of the particle ordering. 

Thus, the denominator term in (\ref{a2g}) is identical to that in (\ref{mhvpartg}) and (\ref{mhvpartq}) for a given particle ordering, while the numerator and colour factors factorize. As it is again the same type of diagrams (corresponding to a given ordering of the external parton momenta) which contributes, we find that the resulting amplitudes will be equivalent up to these overall factors, as expected\footnote{In fact, in this $gg\to q\overline{q} q\overline{q}$ case, the situation is somewhat more complicated. In particular, as we do not include diagrams corresponding to permutations of the legs within each meson state, $l_3(l_5) \leftrightarrow l_4(l_6)$, the singularities discussed above do not cancel and so the amplitudes $A_{a,b,c}^{2g}$ (taking the notation of (\ref{ta})--(\ref{tc})) are not individually finite. However, when all three amplitudes are summed the result is finite. This cancellation of the individual singularities in the three terms relies crucially on the additional factor of $(-1)^p$ which is present in the colour factor of $A_c^{2g}$ (where $p=1$).}.

\bibliography{ggbib}{}
\bibliographystyle{h-physrev}

\end{document}